\documentclass[preprint]{aastex}
\slugcomment{{\sc Accepted to the ApJ:} February 3, 2013}

\usepackage{amsmath}
\usepackage{graphicx}
\usepackage{hyperref}
\usepackage{subfig}
\usepackage{natbib}
\usepackage{longtable}
\usepackage{multirow}
\usepackage{xspace}
\usepackage{pdflscape} 

\newcommand{\snr}{{\raise0.7ex\hbox{${\rm{S}}$} \!\mathord{\left/
 {\vphantom {{\rm{S}} {\rm{N}}}}\right.\kern-\nulldelimiterspace}
\!\lower0.7ex\hbox{${\rm{N}}$}}
}

\newcommand{\Kepler}{\textit{Kepler}\xspace} 

\begin{document}

\author{Andrew P.V. Siemion\altaffilmark{1}, Paul Demorest\altaffilmark{2}, Eric Korpela\altaffilmark{1}, Ron J. Maddalena\altaffilmark{2}, Dan Werthimer\altaffilmark{1}, Jeff Cobb\altaffilmark{1}, Andrew W. Howard\altaffilmark{3}, Glen Langston\altaffilmark{2}, Matt Lebofsky\altaffilmark{1}, Geoffrey W. Marcy\altaffilmark{1}, Jill Tarter\altaffilmark{4}}

\altaffiltext{1}{University of California, Berkeley}
\altaffiltext{2}{National Radio Astronomy Observatory}
\altaffiltext{3}{Institute for Astronomy, University of Hawaii}
\altaffiltext{4}{SETI Institute}

\pagenumbering{roman}
\title{A 1.1 to 1.9 GHz SETI Survey of the \Kepler Field: I. A Search for Narrow-band Emission from Select Targets}
\maketitle
\setcounter{page}{1}
\pagenumbering{arabic}

\section*{Abstract}
We present a targeted search for narrow-band ($< 5 \rm{\ Hz}$) drifting sinusoidal radio emission from 86 stars in the \Kepler field hosting confirmed or candidate exoplanets.  Radio emission less than $5 \rm{\ Hz}$ in spectral extent is currently known to only arise from artificial sources.  The stars searched were chosen based on the properties of their putative exoplanets, including stars hosting candidates with $380\rm{\ K} > T_{\rm{eq}} > 230\rm{\ K}$, stars with 5 or more detected candidates or stars with a super-Earth ($R_{\rm{p}} < 3\ R_{\oplus}$) in a $> 50 $ day orbit.  Baseband voltage data across the entire band between 1.1 and 1.9 GHz were recorded at the Robert C. Byrd Green Bank Telescope between Feb--Apr 2011 and subsequently searched offline.  No signals of extraterrestrial origin were found.  We estimate that fewer than $\sim$1\% of transiting exoplanet systems host technological civilizations that are radio loud in narrow-band emission between 1$-$2 GHz at an equivalent isotropically radiated power (EIRP) of $\sim1.5\times10^{21}{\rm\ erg}{\rm\ s}^{-1}$, approximately eight times the peak EIRP of the Arecibo Planetary Radar, and we limit the the number of 1$-$2 GHz narrow-band-radio-loud Kardashev type II civilizations in the Milky Way to be $<10^{-6}M^{-1}_ \odot$.  Here we describe our observations, data reduction procedures and results.

\section{Introduction}

In the last 50 years, evidence has steadily mounted that the constituents and conditions we believe necessary for life
are common and perhaps ubiquitous in the nearby galaxy.  A plethora of prebiotic molecules have now been detected in molecular clouds, 
including amino acids and their precursors \citep{Mehringer:1997p2145, Kuan:2003eq}, sugars \citep{Hollis:2004p2650} and a host of other biologically important species (e.g. \citealp{Lovas:2006p2493, 2011MNRAS.411.1857I}).  Such detections
offer an indication that the reactants necessary for building large complex organic structures may be formed readily in 
proto-planetary environs.  Exoplanets themselves, while once relegated to the domain of speculation, now appear to be 
common and numerous.  While strictly Earth-size exoplanets at 1 AU from their Solar-type parents have so far eluded detection, we have evidence that the
conditions necessary to maintain carbon-based life, e.g. liquid water, can exist far away from the traditional ``habitable zone'' \citep{Carr:1998p2839}. 

As yet, no evidence exists for the presence of any kind of life outside of the Earth.  However, on our own planet, life is known to have arisen early 
(within 1 Gyr) and flourished \citep{Schopf:2002p2663}.  And while the propensity for evolution of intelligence from basic 
forms of life is not currently well understood, it appears that intelligence has imparted a strong evolutionary advantage 
to our own species.  From a Copernican standpoint, the possibility that life has arisen elsewhere and perhaps evolved intelligence 
is plausible and warrants scientific inquiry.   ``Are we alone as technologically-capable intelligent beings?'' is among the most profound questions we can ask as scientists, and observational astronomy represents the best means of determining an answer.
\vspace{-20px}

\subsection{Engineered Radio Emission}
For a better part of the last century, human beings have produced radio emissions that could readily be recognized as having come from 
no known natural source if transmitted at sufficient power from another star and received on Earth.  These emissions include spectrally
narrow signals, e.g. the sinusoidal carrier waves associated with frequency modulated or amplitude modulated telecommunications,
as well as temporally narrow radio pulses used for radar.  Long wavelength radio photons are efficient and effective interstellar information carriers, as they are energetically cheap and the interstellar medium (ISM) is relatively transparent at radio wavelengths.  The frequency band between $\sim$500 MHz and 10 GHz, the so-called ``terrestrial microwave window'' \citep{Morrison:1977p182} is especially attractive for terrestrial transmission or reception, in that it represents a relatively quiet region of spectrum between the Galactic synchrotron-dominated low frequency spectrum and atmospheric $\rm{H_{2}O}$ and $\rm{O_{2}}$ emission and absorption.   

Natural astrophysical electromagnetic emissions are inherently spectrally broadened by the random processes underlying natural emission physics, with the spectrally narrowest known natural sources, astrophysical masers, having a minimum frequency spread of $\sim$500 Hz \citep{Cohen:1987p3030}. Emission no more than a few Hz in spectral width is, as far as we know, an unmistakable indicator of engineering by an intelligent civilization.  While scintillation effects can render an intrinsically amplitude-stable narrow-band signal intermittent \citep{1997ApJ...487..782C}, narrow-band signals are readily distinguished from background sources of radio emission and are immune to the dispersive effects of the interstellar medium.  Broadband pulsed radio emissions are more deleteriously affected by the ISM, as evidenced by decades of pulsar research, but they too are easily distinguished from incoherent emission and the few thousand known pulsars do not represent a significant interfering background.  Further, some have suggested that directing large amounts of energy into broadband pulsed emission might be attractive than a narrow band transmitter for an economical advanced civilization. \citep{Benford:2008p611}.  

Although the technologies associated with engineered radio emissions from Earth are developed by humans, similar signal types may be used by extraterrestrial intelligent civilizations if they similarly use electromagnetic radiation for ranging and communication.  It is difficult to predict the specific properties of electromagnetic emission from extraterrestrial technologies, but if an extraterrestrial civilization is intentionally indicating its presence via such emission, it would be beneficial to make the signal discriminable.  In terms of distinguishability, both pulsed signals and narrow band signals possess merit, and it is prudent to search for both.  Extrapolating from humanity's exploration of space, it is likely that a more advanced civilization having similar proclivities would explore and perhaps colonize multiple planets in their star system.  These explorations could very easily include planet-planet communication and radar imaging or radar mapping of orbital debris.  Observing planetary systems in which the orbital plane is seen edge-on, such as those identified by transiting exoplanet surveys, thus present a particularly advantageous geometry for eavesdropping on planet-planet electromagnetic signaling by advanced life.

\section{Observations}
\label{sec:obs}
SETI observations were performed during the period February 2011 - April 2011 using the Robert C. Byrd Green Bank Telescope (GBT) L-band (1.1 - 1.9 GHz) receiver and the Green Bank Ultimate Pulsar Processor (GUPPI) digital backend \citep{Demorest:2012vx}.  For this experiment, GUPPI was configured in a novel ``baseband recording'' mode in which an entire 800 MHz band is digitized, channelized to 3.125 MHz with a 256-point polyphase filterbank and written to disk as 2-bit voltage data for both X and Y linear polarizations.  The total aggregate data rate in this mode is 800 MBps.  In order to obtain the requisite disk recording rates, the data stream was distributed to GUPPI's eight CPU/GPU computing nodes at 32 channels/node and eight 100 MHz bands were recorded separately.  For the purposes of analysis, we have considered each 100 MHz band individually, hereafter Bands 0$-$7.  We use this convention primarily because early technical problems with the GUPPI backend resulted in some computing nodes failing to record data, thus causing some observations to have non-contiguous frequency coverage in 100 MHz increments (see Table \ref{tab:obs}).  

Targeted observations were performed on 86 \Kepler Objects of Interest (KOIs) hosting planet candidates judged to be most amenable to the presence of Earth-like life, primarily judged by equilibrium temperature, but also cursory similarity to the Earth and the Solar system.  These targets comprised KOIs hosting planet candidates in or near the traditional ``habitable zone'' ($380\rm{\ K} > T_{\rm{eq}} > 230\rm{\ K}$) as described in \cite{1993Icar..101..108K}, all KOIs hosting 5 or more planet candidates and all KOIs hosting a super-Earth ($R_{\rm{p}} < 3\ R_{\oplus}$) in a $> 50 $ day orbit.  Equilibrium temperatures were taken from \cite{Borucki:2011p9704}, and the extended habitable zone range included in our search reflects their quoted uncertainty of approximately 22\%.  Temperatures were calculated assuming a Bond albedo, emissivity of 0.9 and a uniform surface temperature (see \citealp{Borucki:2011p9704}).  An additional 19 KOIs located within a half power beam width of a primary target were observed serendipitously.  Observations were performed using a cadence in which each target was observed interleaved with another target  separated by $\Theta_{min}>1^{\circ}$ such that each target was effectively observed with an on-source, off-source, on-source sequence with minimal overhead.  This technique is crucial for discerning a true astronomical signal from ubiquitous interference from human technologies.  Details of the parameters of our observations are presented in Table \ref{tab:meters}.  The observations discussed here represent part of a larger 24 hour campaign to search for technologically produced radio emissions in the \Kepler field, which included both targeted observations and a raster scan of the entire field.  Additional work, in preparation, will discuss a narrow-band search of the raster scan data and pulse searches over both targeted and raster observations.    

\begin{table}[h!]
\begin{minipage}{3.2in}
\footnotesize
\caption{Targeted Observation Parameters}
\centering
\vspace{0.1in}
\begin{tabular}{lcl}
\hline\hline\\[-0.10in]
\textbf{Center Frequency} & $\nu_o$ & 1500 MHz \\
\hspace{0.2in}\\[-0.1in]
%\textbf{Number of Channels} & $N_{\textrm{chan}}$ & 128 \\
%\hspace{0.2in}\\[-0.1in]
%\textbf{Channel Width} & $\Delta\nu_i$ & 1.6 MHz \\
%\hspace{0.2in}\\[-0.1in]
\textbf{Bandwidth} & $\Delta\nu$ & 800 MHz \footnote{Excluding the band 1.2 to 1.33 GHz, see main text}\\
\hspace{0.2in}\\[-0.1in]
\textbf{Beam Width} & $\Theta$ & $9'$ \\
(HPBW) & & \\
\hspace{0.2in}\\[-0.1in]
\textbf{System Temperature}\footnote{Nominal value} & $T_{\rm sys}$ & 20 K\\
\hspace{0.2in}\\[-0.1in]
\textbf{Gain}$^{\rm b}$ & $G$ & $2.0$ K/Jy \\
\hspace{0.2in}\\[-0.1in]
\textbf{SEFD}$^{\rm b}$ & $S_{\rm{sys}}$ & $10$ Jy \\
\hspace{0.2in}\\[-0.1in]
\textbf{Observation Time per Source}$^{\rm b}$ & $t_{\rm{obs}}$ & $300$ s \\
\
\end{tabular}
\label{tab:meters}
\end{minipage}
\end{table}

%theta = 2 * pi * (1 - cos(theta/2)) = 0.0176714562 square degrees
%field center 19.3778 44.5027

%stripe centers
%J2000 19.0603 49.2054
%J2000 19.226 46.8799
%J2000 19.3778 44.5027
%J2000 19.5176 42.0858
%J2000 19.6471 39.6317

%Input: Equatorial J2000.0

%285.90450000      49.20540000       0.000000     
%19h03m37.08000s   +49d12m19.4400s
%79.51839036       18.31976923       69.419431

%288.39000000      46.87990000       0.000000     
%19h13m33.60000s   +46d52m47.6400s
%77.90291748       15.91474841       67.094990    

% 290.66700000      44.50270000       0.000000     
%19h22m40.08000s   +44d30m09.7200s
%76.32356640       13.49688473       65.064108    

\section{Data Reduction}
To maximize the signal$-$to$-$noise of the detection of a distant continuous-wave transmitter the relative motion between the transmitter and receiver must be accounted for.  As we have no a priori knowledge of the specific frequency of emission from an extraterrestrial technology, the overall Doppler shift in the received signal, dominated by the radial velocity of the source, is relatively unimportant for detection.  However, the time rate of change of the Doppler shift, dominated by the orbital and rotational motions of the transmitter and receiver, must be considered to integrate $\sim$ Hz spectra over many seconds.  The Doppler drift is given simply by 
\begin{equation}
\dot f = \frac{{d\overrightarrow V }}{{dt}}\frac{{f_{\rm rest} }}{c}
\end{equation}
where $\overrightarrow V$ is the line of sight relative velocity between receiver and source, $f_{rest}$ is the rest frequency of the transmitter and $c$ the speed of light.  As a point of reference, the maximum contribution from the Earth's orbital motion at 1 GHz is $\sim  \pm 0.02 \rm{\ Hz\ s}^{-1}$, and from the Earth's rotation is $\sim  - 0.1 \rm{\ Hz\ s}^{-1}$.  If this effect is corrected for in power spectra, the worst-case minimum achievable spectral resolution for terrestrial observations is thus about $\Delta \nu =0.3$ Hz (at 1 GHz).  Channelization to any finer resolution would be ineffective as the received signal would be smeared over several channels.  While the Doppler drift due to the Earth's motion is known, the drift due to possible motion of the transmitter is largely unknown.  For observations where $t_{\rm{obs}} \delta f  > \Delta \nu$, this necessitates searching various Doppler drift rates to achieve a   $\sqrt {t_{\rm{obs}}}$ increase in sensitivity.  As an aside, an arbitrary Doppler drift can be removed exactly in the voltage domain with no loss in sensitivity, via multiplication by an appropriate chirp function \citep{Leigh:1998p12013}, but this technique is computationally infeasible for most blind searches, an exception being SETI@home \citep{Korpela:2002p3225}.

\begin{figure}[htb]
\includegraphics[width=0.95\textwidth]{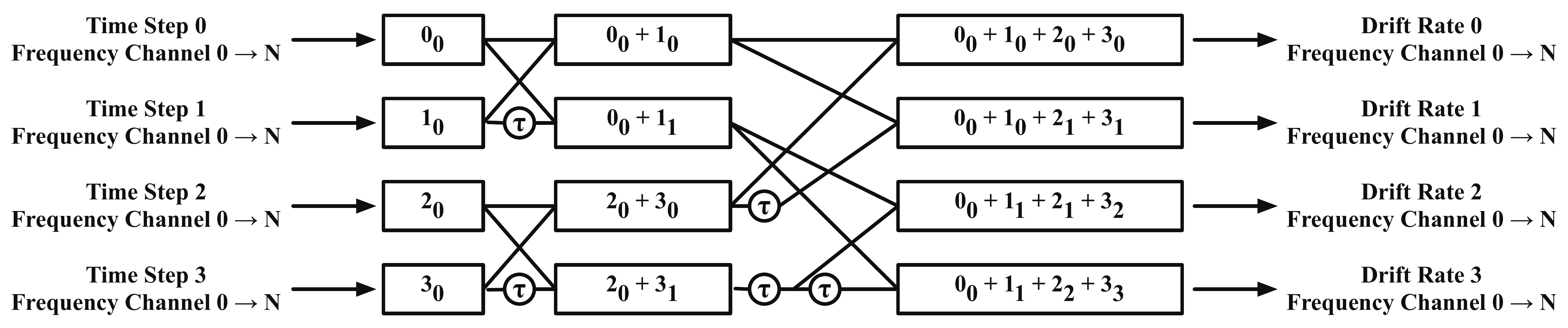}
\caption{
A diagram of the ``tree deDoppler'' algorithm used to search for sinusoids drifting due to Doppler acceleration, shown here for 4 power spectra each having N frequency channels.
\label{fig:dedoppler}
}
\end{figure}

We accomplished a search for narrow-band features drifting at rates up to $\sim  \pm 10 \rm{\ Hz/sec}$ using a modified form of the ``tree'' dedispersion algorithm, an algorithm originally developed for searching for dispersed pulsar emission \citep{1974A&AS...15..367T}.  In much the same way that dispersed pulse searches seek to find power distributed along a quadratic curve in the time$-$frequency plane, a search for drifting sinusoids seeks to find approximately linearly drifting features in the same plane.  The difference is simply one of the dimensions and orientation of the time$-$frequency matrix.  The tree dedispersion algorithm accelerates these searches by taking advantage of the redundant computations involved in searching similar slopes, reducing the number of additions required from $n^2$ to $n {\rm\ log}_2 n$, where n is equal to both the number of spectra and number of slopes searched.  Figure \ref{fig:dedoppler} shows a diagram of the ``tree deDoppler'' algorithm implemented for the drifting sinusoid search, shown here for 4 power spectra each having N frequency channels.  The tree algorithm has fallen into disuse in the pulsar community due to the fact that it intrinsically sums only linear slopes, and modern broadband pulsar observations require a more exact quadratic sum to follow the $\nu^{-2}$ cold plasma dispersion relation.  In the case of Doppler drifting sinusoids, the linear approximation is very good and the tree algorithm is an excellent fit to the problem. 

Our implementation of the ``tree deDoppler'' algorithm necessitates $2^m$ spectra (an integer power of 2) and searches $2^m$ doppler drift rates out to a maximum drift rate of:

\begin{equation}
{\dot f}_{\rm{max}}  = (\Delta \nu )^2 
\end{equation}

where $\Delta \nu$ is the spectral resolution.  The drift rate resolution is constrained to:

\begin{equation}
\Delta \dot f = \frac{{\Delta \nu }}{{T_{\rm{obs}} }}
\end{equation}

where $T_{\rm{obs}}$ is the total observing time.  We performed three channelizations at resolutions ranging from 0.75 Hz to 2.98 Hz, giving the maximum drift rates and drift resolutions shown in Table \ref{tab:driftsearch}.  The maximum drift rate searched would accommodate an equatorial transmitter on a planet 5 times larger and rotating 5 times faster than the Earth.

\begin{table}[h!]
\begin{minipage}{3.8in}
\footnotesize
\caption{Narrowband Search Parameters}
\centering
\vspace{0.1in}
\begin{tabular}{c c c}
\hline\hline\\[-0.10in]
\textbf{Spectral Resolution}& \textbf{Drift Resolution}\footnote{Characteristic value, exact resolution depends on the specific duration of each observation.} & \textbf{Maximum Drift Rate} \\
 (Hz) & (Hz/s) & (Hz/s)\\
\hline
\hspace{0.2in}\\[-0.1in]
2.98 & 0.020 & 8.88 \\
\hspace{0.2in}\\[-0.1in]
1.49 & 0.010 & 2.22 \\
\hspace{0.2in}\\[-0.1in]
0.75 & 0.005 & 0.56 \\
\end{tabular}
\label{tab:driftsearch}
\end{minipage}
\end{table}

Prior to Doppler searching, each 3.125 MHz polyphase channel was further channelized to $\sim$ Hz resolution, detected and thresholded to search for narrow-band features.  After detection, both polarizations were summed to form a single spectrum.  Each high resolution spectrum constructed from individual polyphase channels was corrected for the filter response imposed by first stage (GUPPI) channelization by dividing through with a polynomial fit to an average bandpass.   This polynomial fit bandpass was constructed a priori by fitting to a sum of many coarsely channelized spectra exhibiting low interference (based on visual inspection).  All $M$ spectra having length $N$ from a single observation were then fit into a matrix sized to the nearest larger matrix having dimensions $2^m \times N$.  Because we can assume that an untargetted narrow-band signal transmitted by an extraterrestrial technology will either be drifting due to acceleration in the host system or transmitted uncorrected for the Doppler acceleration at the receiving observatory, any narrow-band signal exhibiting no drift can be ruled out as likely coming from a terrestrial source.  This concept is analogous to searches for pulsars in which sources exhibiting no dispersive sweep in their pulse profiles are likely pulsed terrestrial interference.  The analogy allows us to again borrow from pulsar search techniques and apply a median filter for sources exhibiting no drift (See e.g. \citealp{2009MNRAS.395..410E}, \citealp{2012ApJ...744..109S}).  After dividing each spectral channel by its median value, the tree deDoppler algorithm was applied in-place.  Each Doppler corrected spectrum was then collapsed in time and searched for any summed spectral channel exceeding 25 standard deviations above the mean, assuming Gaussian statistics, with results inserted into a database.  We use the term ``detection'' to refer to one measurement of a unique signal or emitter.  Depending on source intensity, a single signal or emitter can be detected multiple times at different drift rates and bandwidths.  The set of all detections for frequencies $\nu$, standard deviations $\sigma$, drift rates ${\dot f}$ and bandwidths $\Delta\nu$ was searched to identify the detection having the largest $\sigma$ within each spectral window of width ${\dot f}_{\rm{max}}T_{\rm{obs}}$, and a time$-$frequency waterfall plot around this detection was extracted and stored with the corresponding database entry.  We hereafter characterize each of the highest $\sigma$ detections as ``candidate signals.''  Ultimately we were left with approximately $3\times10^{5}$ candidate signals.  

Although the ISM is relatively unobtrusive to narrow-band radio emission, relative to e.g. interstellar dust on optical light, and the atmosphere (including the ionosphere) are essentially transparent between 1--10 GHz, the signals being considered here are not wholly unperturbed by the intervening media.  The ISM has been considered in the context of SETI for some time, notably in \cite{1991ApJ...376..123C}, \cite{1997ApJ...487..782C} and references therein, with the principal results being as follows.  In the strong scattering regime, narrow band sinusoids experience limited spectral broadening due to scattering in the inhomogeneous interstellar plasma, with a bandwidth $ \Delta \nu _{{\rm{broad}}}$ equal to:
\begin{equation}
\label{eq:dnu}
\Delta \nu _{{\rm{broad}}}  = 0.097{\rm{\ Hz\ }}\nu _{{\rm{GHz}}} ^{ - 6/5} {\left( {\frac{{V_ \bot  }}{{100}}} \right)} {\rm{SM}}^{ 3/5} 
\end{equation}
Where $V_{\bot}$ is the transverse velocity of the source in km/sec and SM the scattering measure, a measure of the electron density fluctuations ${C_{n_e } ^2 }$, (c.f. \citealp{1990ARA&A..28..561R}) integrated along the line of sight:
\begin{equation}
{\rm SM} = \int_0^L {C_{n_e } ^2 } (z)dz
\end{equation}
Further, intrinsically amplitude-stable narrow-band emission can be modulated in intensity up to 100\% by strong scattering in the inhomogeneous plasma, with a characteristic time scale $\Delta t_d$ equal to: 
\begin{equation}
\label{eq:dt}
\Delta t_d  = 3.3{\rm{\ s\ }}\nu _{{\rm{GHz}}} ^{6/5} {\rm{ }}{\left( {\frac{{V_ \bot  }}{{100}}} \right)}^{ - 1}  {\rm{SM}}^{ - 3/5} 
\end{equation} 
Taking values of the SM from the ``NE2001'' electron density model \citep{2002astro.ph..7156C} for a center-field \Kepler star at 0.5 kpc and assuming a transverse velocity of 25 km/sec, we calculate $\Delta t_d  \approx  3.5 {\rm\ hrs}$   and $\Delta \nu _{{\rm{broad}}} \approx 20 {\rm\ }\mu{\rm Hz}$.  

We note that the transition from strong to weak ISM scattering for our observing band occurs at a distance of about 900 ly in the direction of the \Kepler field, putting many of our targets in the transition or weak scattering regimes.  Although the expressions for $\Delta \nu$ and $\Delta t_d$ differ for these cases (see e.g. \citealp{1990ARA&A..28..561R}) , the predicted broadening is similarly negligible, intensity modulations significantly lower in amplitude and modulation time scale longer in duration.

Depending on line of sight, the solar wind and interplanetary medium (IPM) can also impose significant spectral broadening on a transiting narrow-band signal.  Radio scintillation due to the IPM had been known for some time prior to the discovery of strong scattering in the ISM, studied primarily through angular broadening of distant compact radio sources (see \citealp{1992RSPTA.341..151N} and references therein).  The presence of spacecraft that could be used as monochromatic and coherent radio test sources allowed an additional probe of the IPM, notably providing a means to measure not just electron density fluctuations but also solar wind velocity (\citealp{1978ApJ...219..727W}, \citealp{1979JGR....84.7288W}), through observations of spectral broadening and phase scintillation of their narrow carrier signals.  The strength of these effects depend largely on the solar impact distance $R$, or line-of-sight solar separation angle, but significant longitudinal and temporal (e.g. the solar cycle and coronal mass ejections) variations occur as well \citep{2009RaSc...44.6004M}.   \cite{2007SpWea...509004W} presents an assimilation of phase scintillation and spectral broadening observations of the S-band (2.3 GHz) carrier on Pioneer and Helios spacecrafts at solar impact distances up to ~200 $R_{\odot}$ (adapted from \citealp{1978ApJ...219..727W}), and find a roughly $R^{-9/5}$ dependence for spectral broadening past $R\sim 10\  R_{\odot}$.  A variety of models and observations suggest that the electron density fluctuations in the solar wind follow an approximate power law density spectrum, with a mean index very close to the Kolmogorov value of $5/3$ (\citealp{2009RaSc...44.6004M} and references therein), allowing these results to be extrapolated based on $\Delta \nu _{{\rm{broad}}}$ scaling as $\nu^{ - 6/5}$.  Scaling based on the results in \cite{2007SpWea...509004W}, we calculate a spectral broadening contribution from the IPM of approximately:

\begin{equation}
\label{eq:dnusolar}
\Delta \nu _{{\rm{broad}}}  = 300 {\rm{\ Hz\ }}\nu _{{\rm{GHz}}} ^{ - 6/5} {\left( {\frac{R}{R_{\odot}}} \right)} ^{-9/5}
\end{equation}

\noindent for solar impact distances greater than $\sim 10\ R_{\odot}$.        

For our observations, the nearest solar impact distance was $\sim195 \ R_{\odot}$, giving an IPM spectral broadening contribution of $\sim14 \ {\rm mHz}$.  

For the parameters of this search, we can thus neglect spectral broadening due to either the ISM or IPM, and can assume a relatively steady flux for any intrinsically continuous  and amplitude-stable signal over the course of our observing cadence.  A key result of \cite{1991ApJ...376..123C} was the suggestion that searches for narrow-band emissions in the strong scattering regime would have an increased likelihood of detecting a source by observing a sky location multiple times spanning many $\Delta t_d$.  Although this is a very well justified strategy, the extra observing overhead associated with performing multiple on$-$off$-$on observation sequences didn't permit it to be used here.

\section{Analysis}  
\label{sec:anal}
The principal complication in the otherwise straightforward data reduction involved in a narrow-band SETI experiment is the fact that human radio technology produces copious narrow band emission at $\sim$ Hz scales.  The existence of radio frequency interference is not unique to SETI experiments, of course, and over the years many techniques have been developed to mitigate its effects.  It is worth noting, however, that in radio observations of most astrophysical phenomena, narrow-band features in a power spectrum can be immediately flagged and discarded because they are known to originate with technology rather than the target of the observation.  Narrow-band radio SETI experiments face the more difficult task of determining whether a narrow-band feature originates with a human technology or distant intelligent life.    

\begin{figure}[htb]
\includegraphics[width=0.65\textwidth]{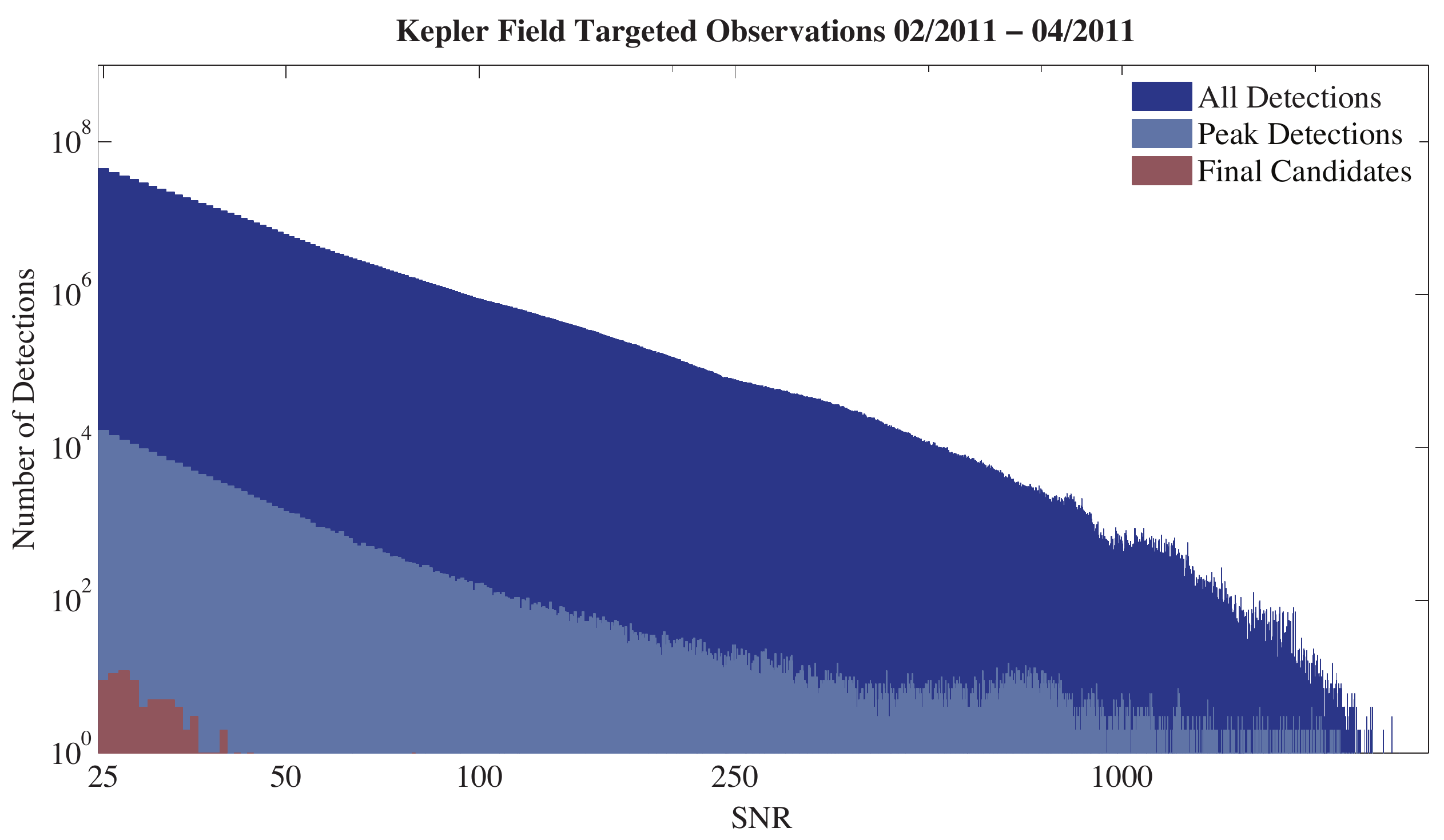}
\caption{
Number of detections vs. signal-to-noise ratio for the set of all detections, the detections representing the most significant detection of a single signal and only those candidates passing an automated interference excision algorithm (See Section \ref{sec:anal}).
\label{fig:snrhist}
}
\end{figure}

\begin{figure}[htb]
\includegraphics[width=0.65\textwidth]{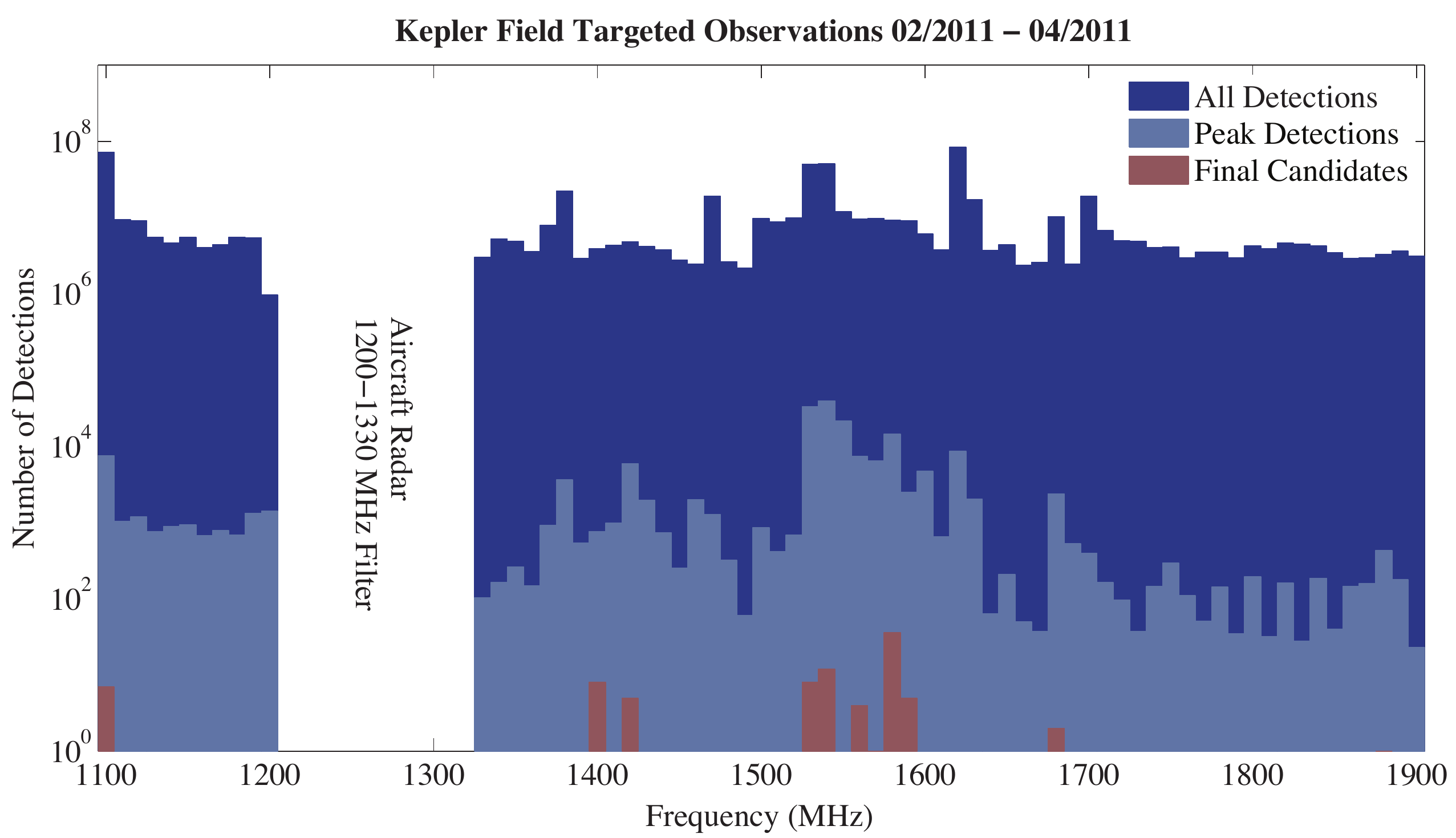}
\caption{
Number of detections vs. topocentric frequency for the set of all detections, the detections representing the most significant detection of a single signal and only those candidates passing an automated interference excision algorithm (See Section \ref{sec:anal}).  A region of spectrum between 1200$-$1330 MHz was excluded due to the presence of a strong interfering radar.
\label{fig:freqhist}
}
\end{figure}

\begin{landscape}
\begin{figure}
    \centering
    \subfloat[]{\label{squiggle:p0}\includegraphics[scale=0.49]{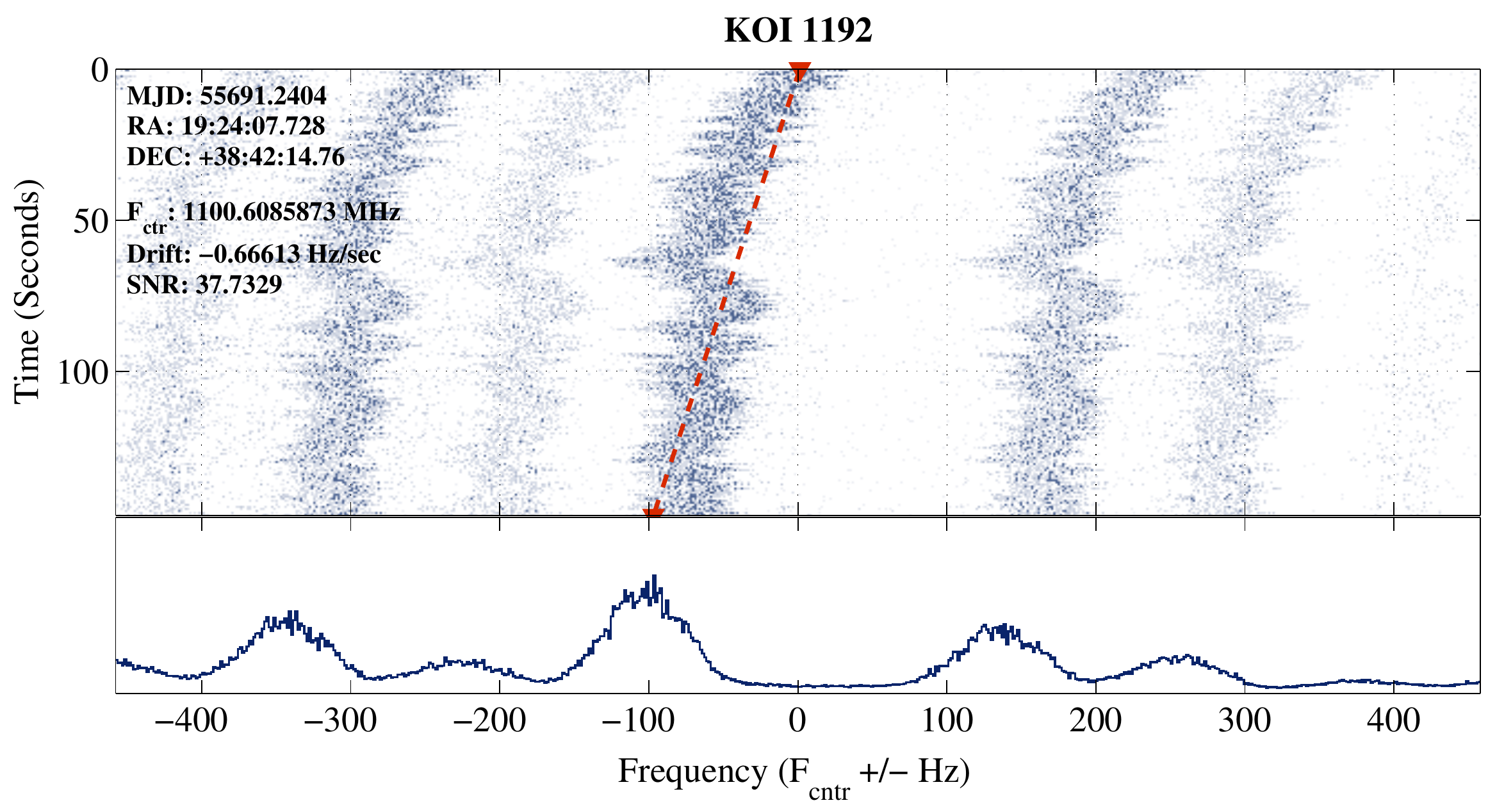}}
    \subfloat[]{\label{squiggle:p1}\includegraphics[scale=0.49]{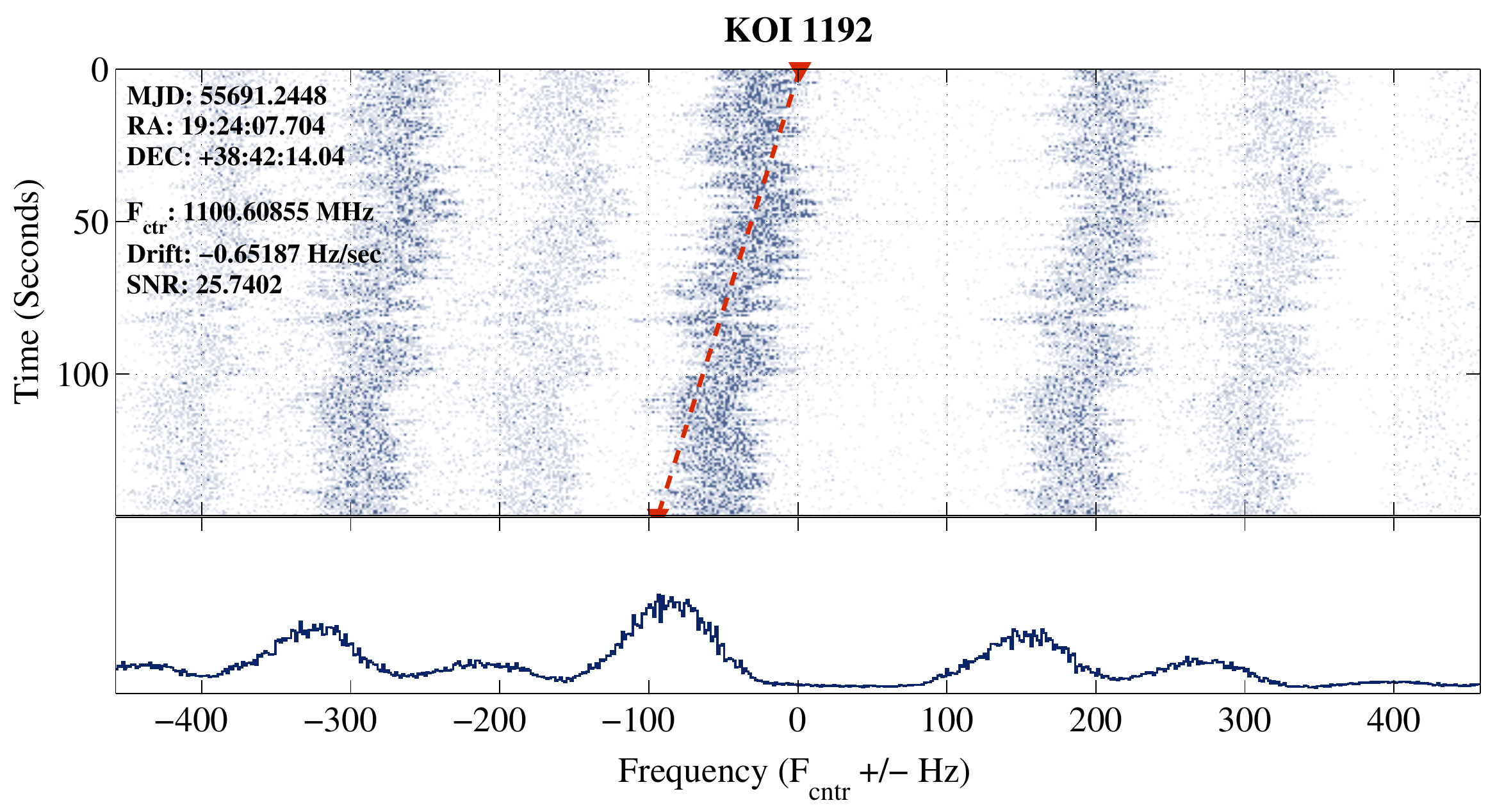}}\\
    \subfloat[]{\label{squiggle:p2}\includegraphics[scale=0.49]{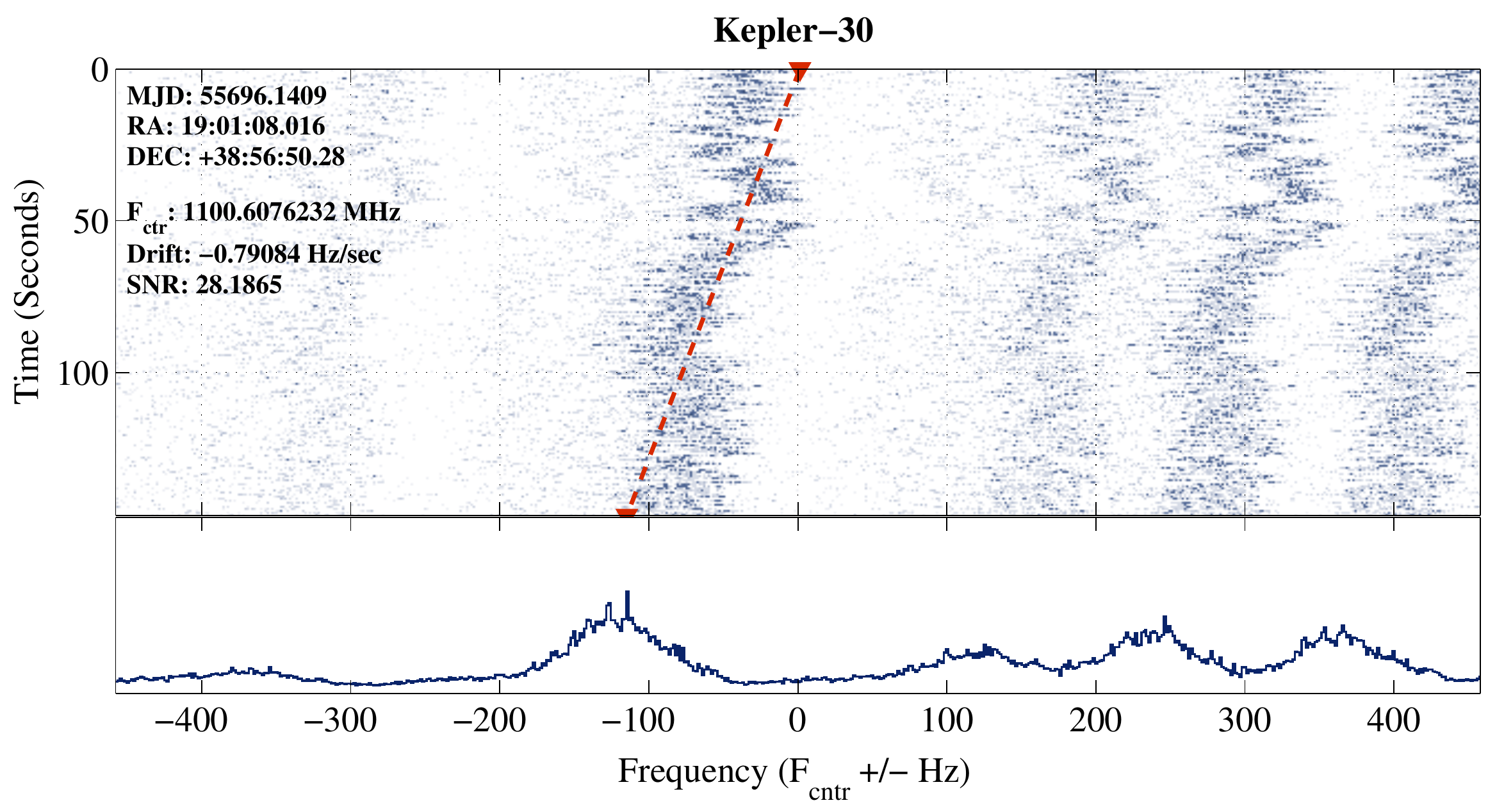}}
    \subfloat[]{\label{squiggle:p3}\includegraphics[scale=0.49]{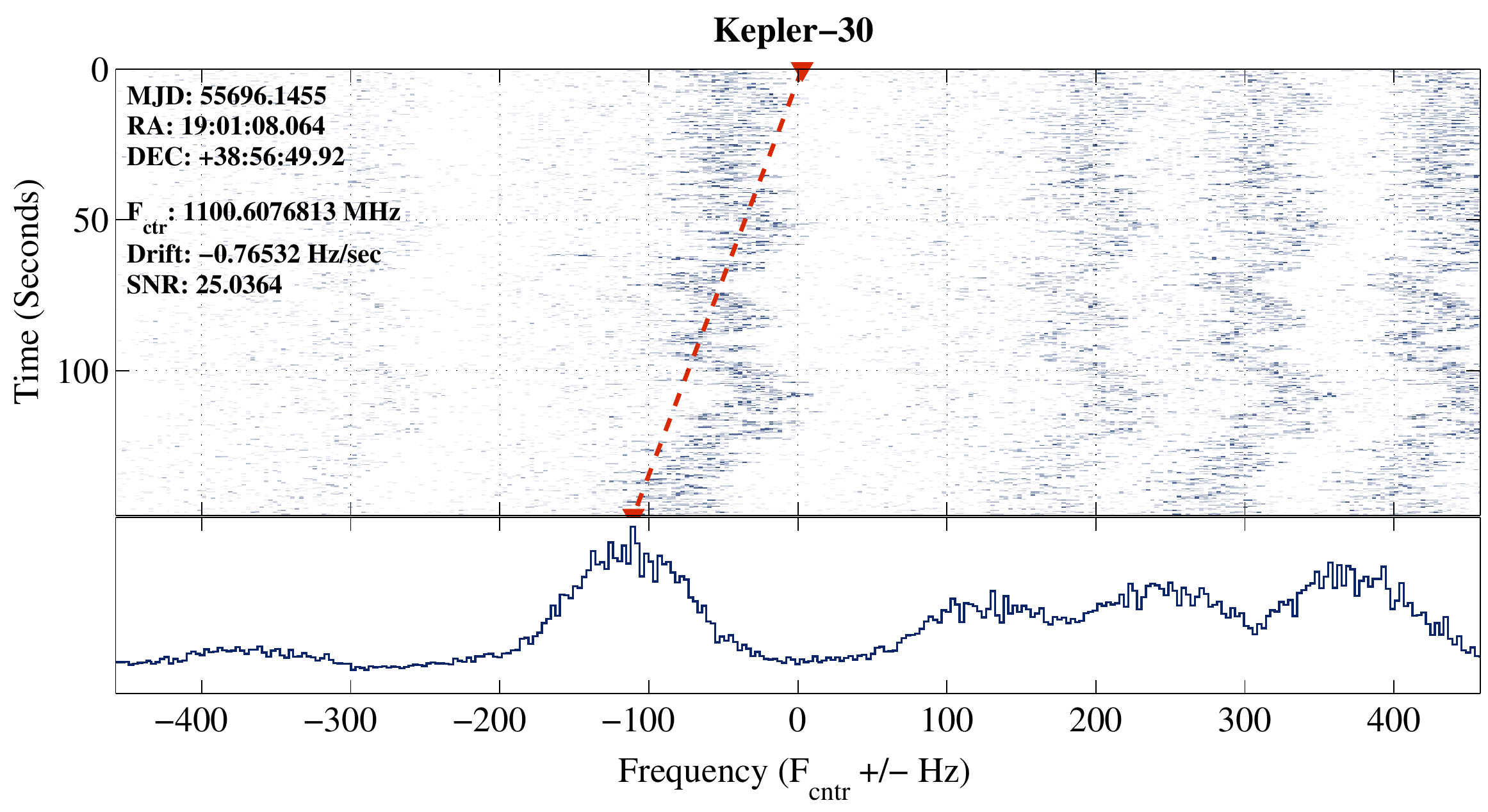}}
    \caption{Waterfall plots showing narrow band emissions, all of which were determined to be interference based on similar topocentric frequency and modulation.  The upper portion of each panel, in blue, shows intensity as a function of topocentric frequency and time and the lower portion of each panel shows a ``Doppler-corrected spectrum'' -- a power spectrum for the entire observation formed by summing consecutive spectra at the drift rate indicated by the red diagonal.  Panels \ref{squiggle:p0} and \ref{squiggle:p1} show detections of the interferer during two pointings on KOI 1192 separated by $\sim$380 s.  Panels \ref{squiggle:p2} and \ref{squiggle:p3} show detections of a very similar interferer approximately 5 days later during two pointings on \Kepler-30. 
    \label{fig:squiggle}}
\end{figure}
\end{landscape}

\begin{landscape}
\begin{figure}
    \centering
    \subfloat[]{\label{gps:p0}\includegraphics[scale=0.49]{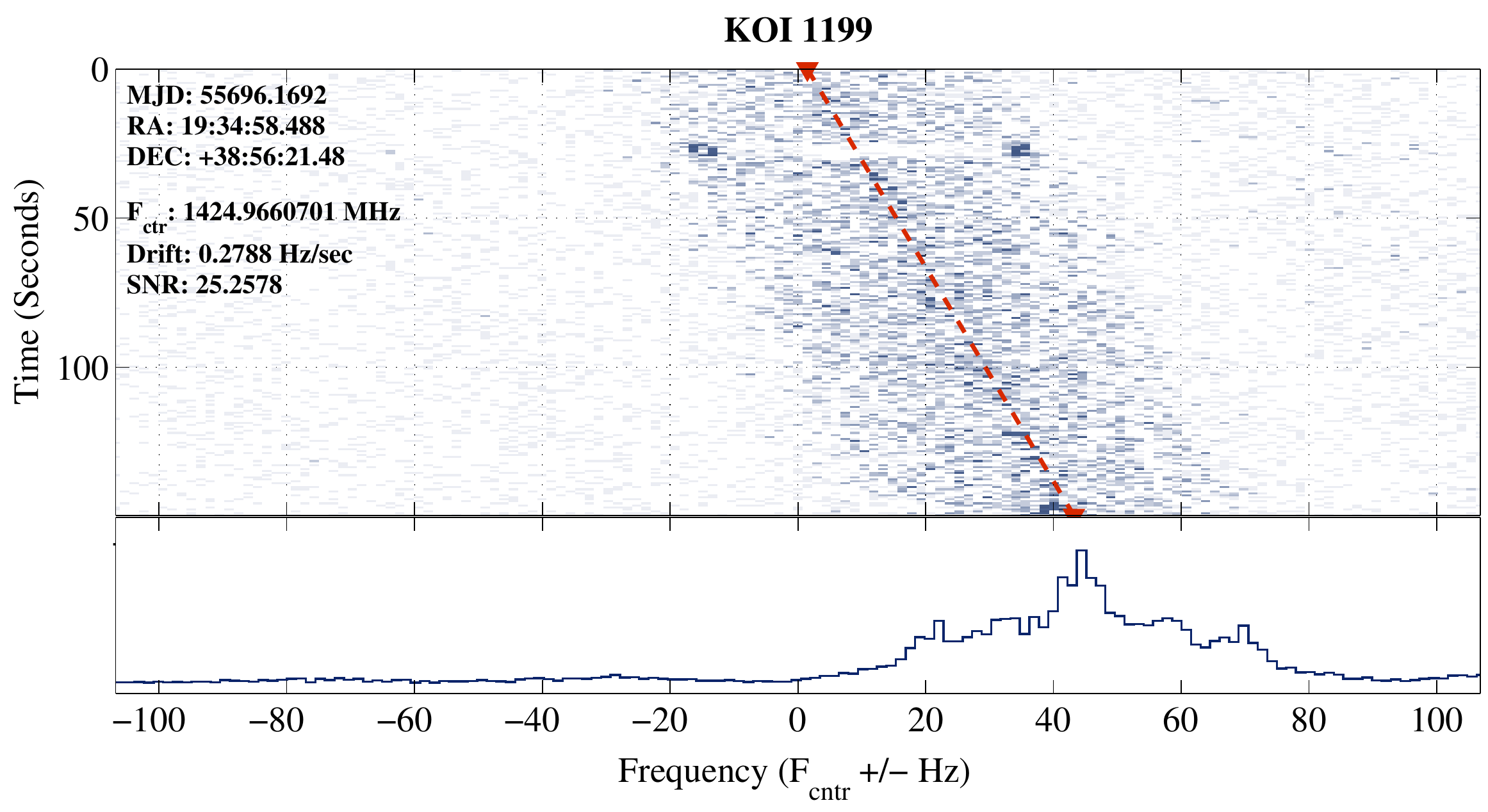}}
    \subfloat[]{\label{gps:p1}\includegraphics[scale=0.49]{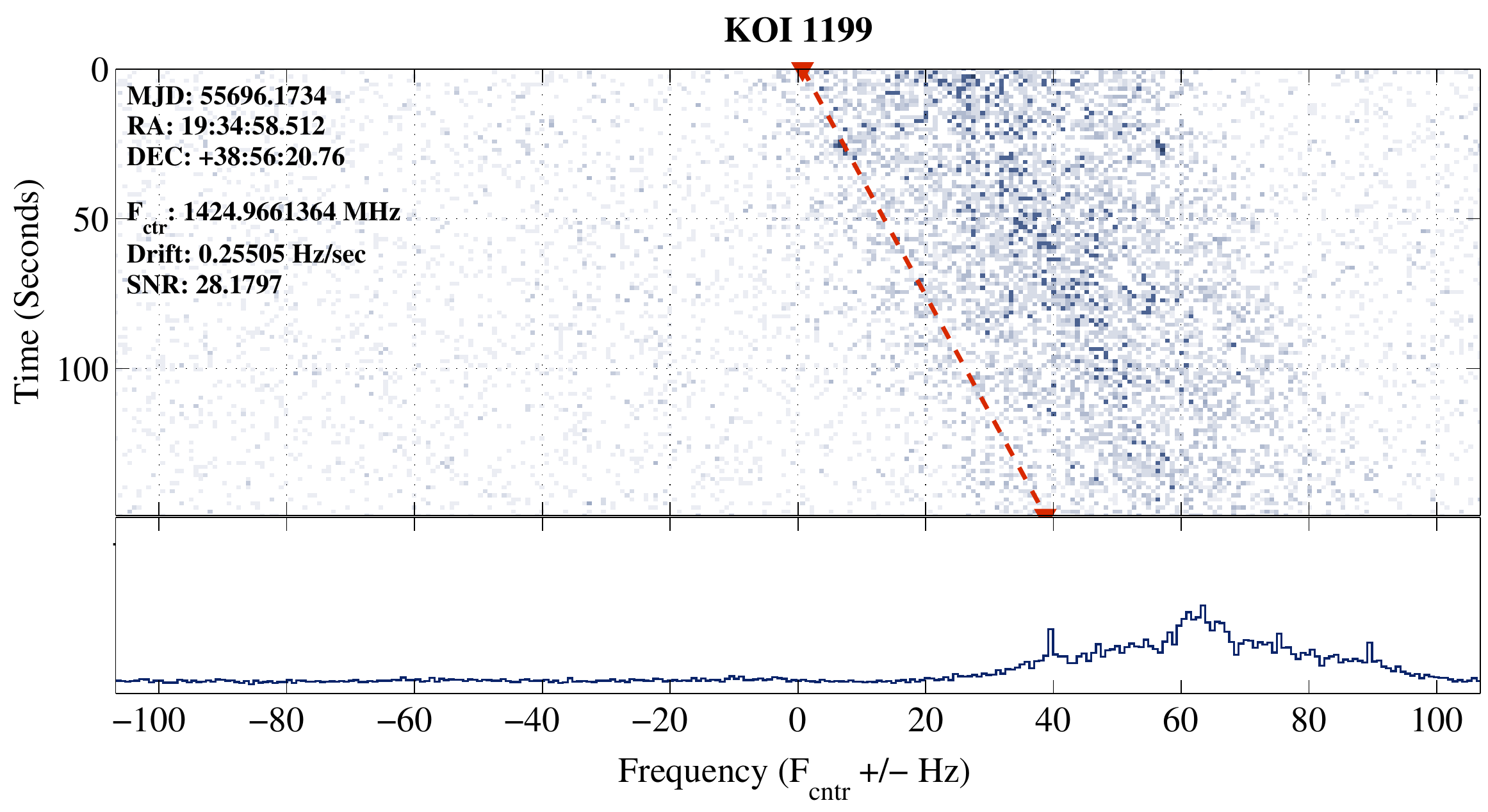}}\\
    \subfloat[]{\label{gps:p2}\includegraphics[scale=0.49]{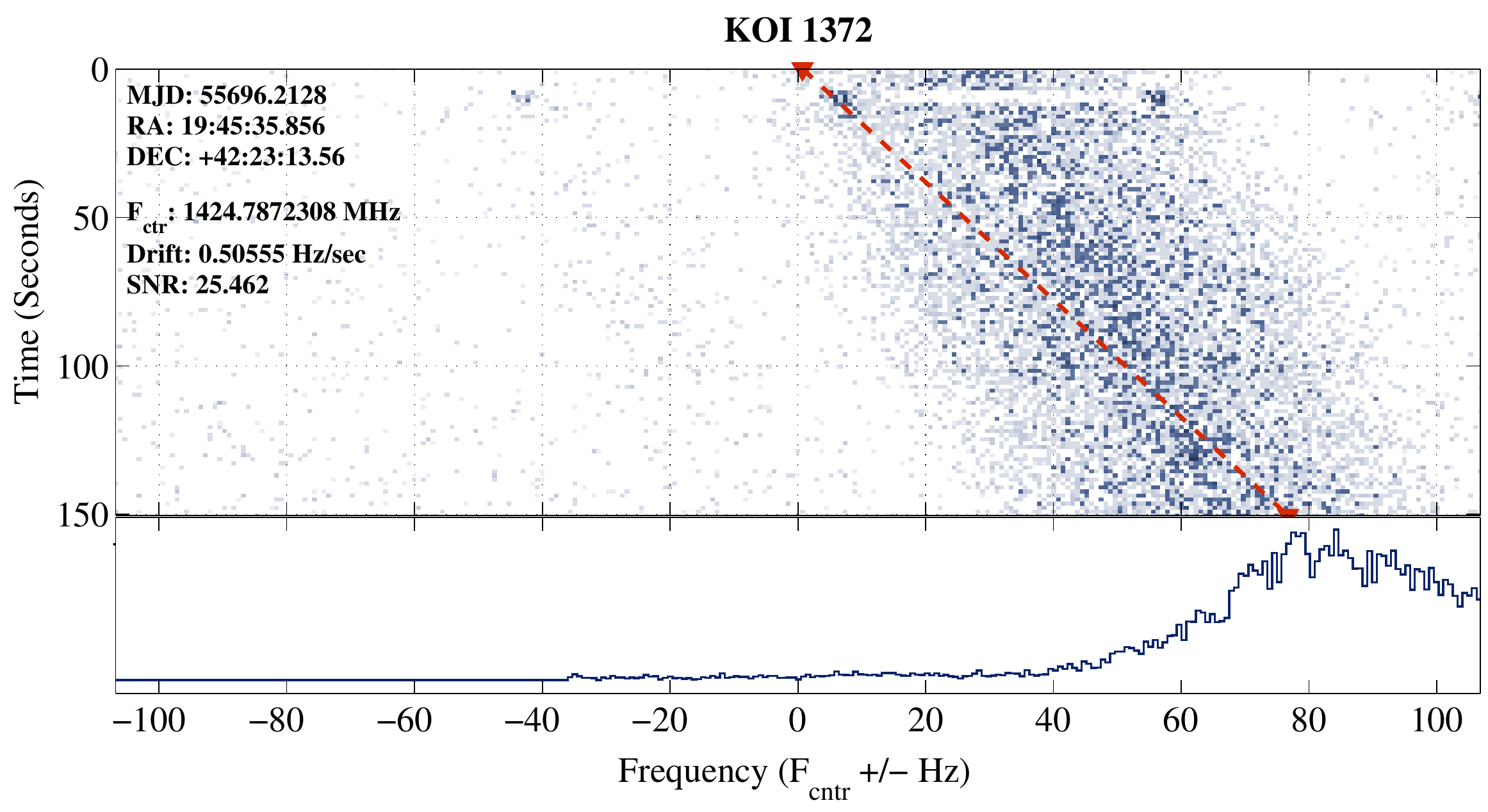}}
    \subfloat[]{\label{gps:p3}\includegraphics[scale=0.49]{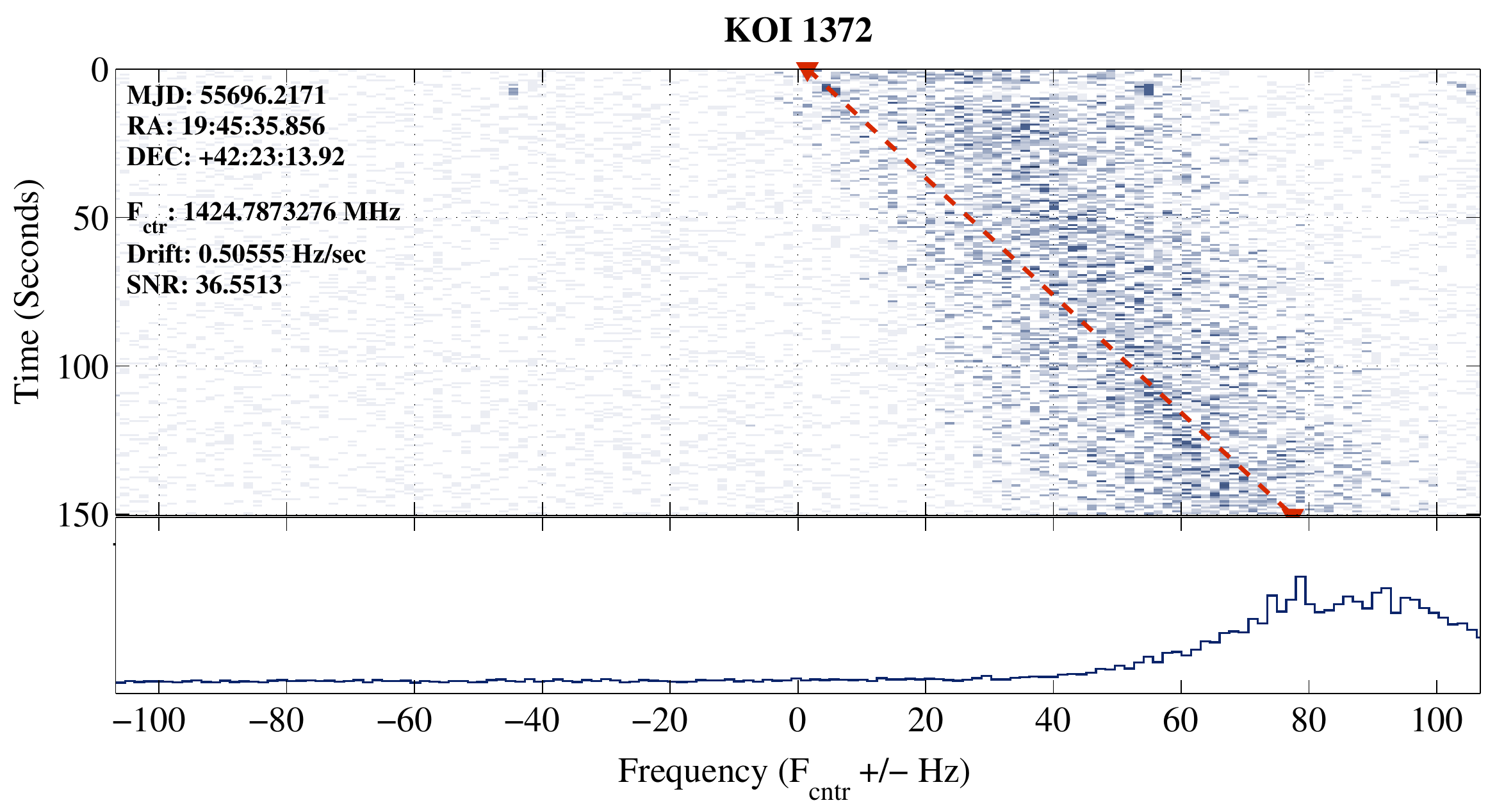}}
    \caption{Waterfall plots showing narrow band emissions, all of which were determined to be interference based on similar bandwidth and modulation.  See Figure \ref{fig:squiggle} caption for plot descriptions.   Panels \ref{gps:p0} and \ref{gps:p1} show detections of the interferer during two pointings on KOI 1199 separated by $\sim$360 s.  Panels \ref{gps:p2} and \ref{gps:p3} show detections of a very similar interferer approximately 1 hour later during two pointings on KOI 1372. 
    \label{fig:gps}}
\end{figure}
\end{landscape}

Our strategy to mitigate terrestrial interference was to demand that a candidate signal be both persistent and isolated on the celestial sphere.  By observing in an on$-$off$-$on  source cadence, we imposed this constraint by requiring that a given candidate signal be detected in both ``on'' source observations and not in the intervening ``off'' source observation.  Observations in which one of the elements of the on$-$off$-$on cadence was not obtained due to technical problems were excluded completely.  This technique was very effective, ruling out 99.96\% of the candidate signals.  Figure \ref{fig:snrhist} shows a histogram of the number of detections vs. signal$-$to$-$noise ratio for all detections, the detections representing the most significant detection of a single emitter (candidate signals) and only those detections passing the on$-$off$-$on automated interference excision algorithm.  Figure \ref{fig:freqhist} shows the number of detected signals as a function of topocentric frequency for the same detection groups.  Time$-$frequency waterfall plots of the remaining 52 candidate signals were examined visually.  Of these, 37 were ruled out immediately because candidates detected during pointings at many targets exhibited very similar modulation at nearby frequencies ($\pm$ 1 MHz).  The remaining 15 signals were ruled out after querying the entire database of candidate signals for detections within $\pm$ 1 MHz and identifying signals detected during pointings at other targets that closely resembled the modulation and drift properties of the candidate signals.  Figure \ref{fig:squiggle} shows two candidate signals that passed our initial on$-$off$-$on  test, but were ruled out as interference based on their similar topocentric frequency and modulation.  Figure \ref{fig:gps} shows two candidate signals detected at different topocentric frequencies, but that were ruled out as interference based on their similar bandwidths and modulation.

\subsection{Sensitivity}

From the radiometer equation, the minimum detectable flux, $F_i$, of narrow-band emission detected in a single polarization is given by\footnote{Assuming the intrinsic received emission width is $<\Delta b$, the spectral channel bandwidth}:

\begin{equation}
F_i  = \sigma _{{\rm{thresh}}} S_{\rm{sys}} \sqrt {\frac{\Delta b}{t}} 
\end{equation}

Where $\sigma _{{\rm{thresh}}}$ is the signal/noise threshold,  $S_{\rm{sys}}$ is the system equivalent flux density (SEFD) of the receiving telescope, $\Delta b$ is the spectral channel bandwidth and $t$ the integration time.  Assuming a flat 10 Jy SEFD for the GBT's L-band receiver, a characteristic sensitivity for the observations presented here is $\sim 2 \times 10^{-23} {\rm\ erg\ } {\rm s }^{-1}{\rm cm}^{-2}$ or $\sim 3$\ Jy across a 0.75 Hz channel.  A useful fiducial for considering the detectability of an extraterrestrial technology at radio wavelengths is the luminosity of the most powerful radio transmitter on Earth, the Arecibo Planetary Radar.  Arecibo hosts two radar systems, a 430 MHz system capable of pulsed operation at $\sim 2.5\ \rm{MW}$ peak for a $\sim 5\%$ duty cycle and a 2380 MHz system producing $\sim 1\ \rm{MW}$ continuous power.  The equivalent isotropically radiated power (EIRP) of the higher frequency system, approaching 20 TW, is the larger of the two, owing to the $\nu^{2}$ gain improvement.  We consider this luminosity value $L_{\rm AO} \approx 2\ \rm{x}\ 10^{20}\ erg\ s^{-1}$ to be approximately equal to the best-case radio emission from an Earth-level technology.  Coincidently, $L_{\rm AO}$ is approximately the same as the current total average power used by all humans on the planet Earth \citep{Gruenspecht:2010tr}.  At 0.5 kpc, a 1 $L_{\rm AO}$  transmitter beamed in the direction of Earth would have a total flux of about $10^{-24} {\rm \ erg\ } {\rm s }^{-1}{\rm cm}^{-2}$, placing our detectable limit for this range at $\sim 8\ L_{\rm AO}$.  Table \ref{tab:obs} details total on-source times and sensitivities for all observed sources. 

\begin{figure}[htb]
\centering
\includegraphics[width=0.75\textwidth]{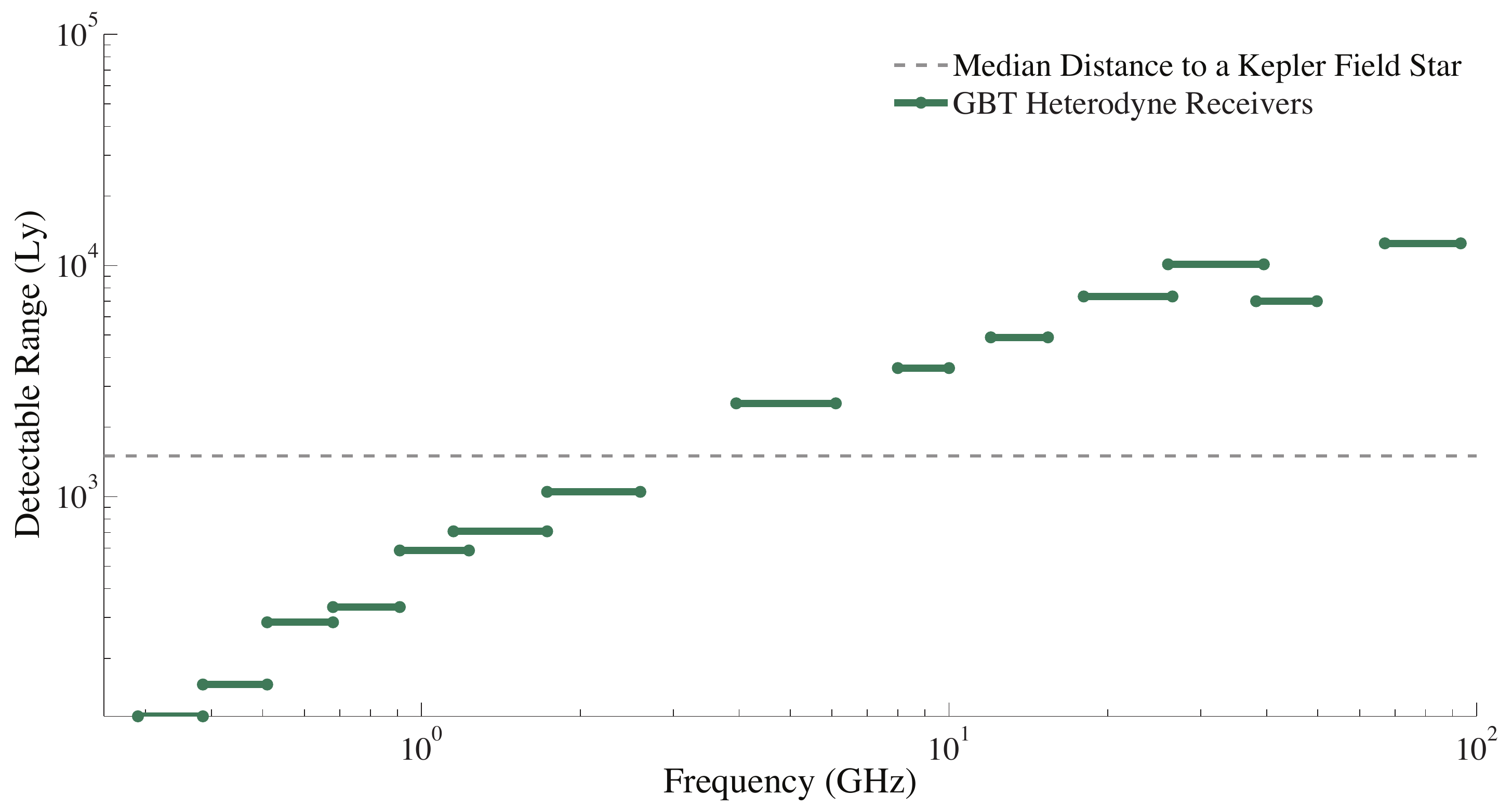}
\caption{The range at which a transmitter similar to the Arecibo Planetary Radar ($\sim 5{\rm\:x\:}10^{13} \: {\rm erg \; s}^{-1}$ transmitted through a 305$\:$m parabolic reflector) could be detected using the parameters of this experiment (150 second integrations, 0.75 Hz channelization) applied to all heterodyne receivers at the GBT.
\label{fig:range}
}
\end{figure}

Figure \ref{fig:range} shows the range at which a transmitter similar to the Arecibo Planetary Radar ($\sim 5{\rm\:x\:}10^{13} \: {\rm erg \; s}^{-1}$ transmitted through a 305$\:$m parabolic reflector) could be detected using the parameters of this experiment (150 second integrations, 0.75 Hz channelization) applied to all heterodyne receivers at the GBT.  These limits also apply to an intrinsically uncertainty-limited broadband pulse having approximately 1000 times the total radiated energy, broadened assuming the ``NE2001'' ISM model [Cordes and Lazio, 2001] to $t=0.17\ \mu\rm{sec}$ with pulse bandwidth = 800 MHz centered on our observing band.  Here we have used system temperature and gain values from the GBT Proposer's Guide, neglected the galactic synchrotron background and for frequencies above 15 GHz we assume a 50\% weather quantile.  We have assumed a 25 $\sigma$ detection threshold, as was used in the analysis described here.  The approximate median distance to a \Kepler catalog star is also indicated.  Although Figure \ref{fig:range} suggests higher frequencies might be preferred, again owing to the transmitter gain improvement, the additional scheduling difficulties due to weather constraints and pointing correction overhead make lower frequency observations more tractable at present.   

We use the Arecibo Planetary Radar example simply as a point of reference.  While this transmitter is highly directional and the probability of interception is thus fairly low, observing systems in which the ecliptic plane is viewed edge-on increases the probability of detecting a radar used for local planetary system ranging and imaging.  If we assume that an exo-Arecibo has a duty cycle similar to Earth's, a characteristic ecliptic ($\pm 5^{\circ}$) illumination of about 100 hours/year \citep{Perillat:2012ve}, the overall probability of being in the radar beam during an observation is $\sim 2 \times 10^{-8}$. While this figure is indeed low, it represents an order of magnitude improvement over an isotropic assumption.

\section{Discussion and Summary}
Our search of 104 KOIs identified no evidence of advanced technology indicative of intelligent life.  If we assume a low false positive rate for KOIs, in the simplest terms this result indicates that fewer than $\sim$1\% of transiting exoplanet systems are radio loud in narrow-band emission between 1$-$2 GHz at the $\sim 8\ L_{\rm AO}$ level.  If we take the orbital inclination requirement for detection to be $\pm 5^{\circ}$ and conservatively estimate the total fraction of FGK stars hosting planets of any type to be $\sim$15\% \citep{2011IAUS..276....3M}, we estimate that fewer than $\sim10^{-4}$ FGK stars host civilizations detectable via orbital plane narrow-band radio emission in the same band and luminosity.  For the GBT, this implies a surface density of 
$<\sim5\times10^{-2}{\rm\ deg^{-2}}$ detectable sources . For the upcoming Square Kilometer Array, a facility that will be perhaps 100 times more sensitive than the GBT, the similar surface density of detectable sources is $<\sim100{\rm\ deg^{-2}}$ ($l=0,b=0$; \citealp{2003A&A...409..523R}). 

Although the observations described here were part of a campaign targeting specific KOIs, the size of the telescope beam probed a much larger population of stars at a concomitantly higher luminosity limit.  When probing advanced technology luminosities that represent a reasonable extrapolation of terrestrial technology, i.e. $\approx$ Kardashev type I \citep{1964SvA.....8..217K}, describing the target population and quantifying limits based on number of Sun-like stars or number of Earth-like planets is quite logical.  However, when we begin to probe luminosities (and energy usage) that are many orders of magnitude larger than the Earth's, our uncertainty in the bounds of life in general render these measures inadequate.  For large luminosities, total stellar mass is a much more useful measure of the amount of energy-delivering capacity of a surveyed area.  Integrating the GBT's beam out to and encompassing the Milky Way's halo stars \citep{Gnedin:2010vr}, a characteristic total mass is $\sim 5\times10^{3}\ M_ \odot$.  At 80 kpc, our sensitivity equates to an EIRP limit of $\sim 10^5\ L_{\rm AO}$, or approximately an order of magnitude larger than the total solar insolation incident on Earth.  A civilization capable of truly isotropic emission at these power levels would likely be capable of harnessing vastly greater amounts of energy from their parent sun than incident on their home planet, and thus would be approaching the Kardashev type II class.  Taking our 86 observations as independent samples of a $\sim 5\times10^{3}\ M_ \odot$ column, we estimate the number of 1$-$2 GHz narrow-band-radio-loud Kardashev type II civilizations in the Milky Way to be $<10^{-6}\ M^{-1}_ \odot$.

%Our search targeted 
                         
%field center J2000 19.3778 44.5027
%                         76.32356640       13.49688473
 %                        76.3236¡  13.4967¡
                                  
%0.00002153212
%$\sim 2 \times 10^{-23} {\rm\ erg\ } {\rm s }^{-1}{\rm cm}^{-2}$
%total stellar mass $5* 10 ^3$

Ultimately, experiments such as the one described here seek to firmly determine the number of other intelligent, communicative civilizations outside of Earth.  However, in placing limits on the presence of intelligent life in the galaxy, we must very carefully qualify our limits with respect to the limitations of our experiment.  In particular, we can offer no argument that an advanced, intelligent civilization {\it necessarily} produces narrow-band radio emission, either intentional or otherwise.  Thus we are probing only a potential subset of such civilizations, where the size of the subset is difficult to estimate.  The search for extraterrestrial intelligence is still in its infancy, and there is much parameter space left to explore.  The exponential growth in semiconductor technology over the last decades has been an incredible boon to SETI experiments, allowing orders of magnitude improvements in spectral coverage.  Within the next decade, we will have the ability to examine significantly larger portions of the electromagnetic spectrum, including instantaneous analysis of the entire 10 GHz of the terrestrial microwave window.  In addition to radio searches, new technology will extend SETI into regions of the electromagnetic spectrum never before observed with high sensitivity \citep{2011arXiv1109.1136S}.  Extending searches to encompass much larger classes of signals is crucial to producing robust and meaningful limits.

%scintime = 3.3 * nu**1.2 * sm**(-0.6) * (100./vperp)
%specbroad = 0.097 * nu**(-1.2) * sm**0.6 * (vperp/100.)	! Hz
%Average transverse velocity: ~30km/sec

\section{Acknowledgements}
We thank John Ford and Scott Ransom for technical assistance during our observations and Gerry Harp for comments on an early draft of this manuscript.  The work presented here was partially funded by NASA Exobiology Grant NNX09AN69G and donations from the Friends of Berkeley SETI and the Friends of
SETI@home.  We also acknowledge the financial and intellectual contributions of the students, faculty and sponsors of the
Berkeley Wireless Research Center.  This research used resources of the National Energy Research
Scientific Computing Center, which is supported by the Office of
Science of the U.S. Department of Energy under Contract No. 
DE-AC02-05CH1123.    APVS gratefully acknowledges receipt of student observing support from the 
National Radio Astronomy Observatory.  The National Radio Astronomy Observatory is a facility of the National Science Foundation operated under cooperative agreement by Associated Universities, Inc.
\clearpage

\bibliography{references}

\begin{thebibliography}{34}
\expandafter\ifx\csname natexlab\endcsname\relax\def\natexlab#1{#1}\fi

\bibitem[{Benford {et~al.}(2008)Benford, Benford, \&
  Benford}]{Benford:2008p611}
Benford, J., Benford, G., \& Benford, D. 2008, arXiv:physics.pop-ph/0810.3964v2

\bibitem[{Borucki {et~al.}(2011)Borucki, Koch, \& Team}]{Borucki:2011p9704}
Borucki, W.~J., Koch, D.~G., \& Team, K.~S. 2011, arXiv, astro-ph.EP
  1102.0541v1

\bibitem[{Carr {et~al.}(1998)Carr, Belton, Chapman, \& Davies}]{Carr:1998p2839}
Carr, M., Belton, M., Chapman, C., \& Davies, M. 1998, Nature, 391, 363

\bibitem[{Cohen {et~al.}(1987)Cohen, Downs, Emerson, Grimm, Gulkis, Stevens, \&
  Tarter}]{Cohen:1987p3030}
Cohen, R., Downs, G., Emerson, R., Grimm, M., Gulkis, S., Stevens, G., \&
  Tarter, J. 1987, Mon. Not. R. Astron. Soc.

\bibitem[{Cordes \& Lazio(1991)}]{1991ApJ...376..123C}
Cordes, J.~M., \& Lazio, T.~J. 1991, Astrophysical Journal, 376, 123

\bibitem[{Cordes \& Lazio(2002)}]{2002astro.ph..7156C}
Cordes, J.~M., \& Lazio, T. J.~W. 2002, arXiv:0207156v3

\bibitem[{Cordes {et~al.}(1997)Cordes, Lazio, \& Sagan}]{1997ApJ...487..782C}
Cordes, J.~M., Lazio, T. J.~W., \& Sagan, C. 1997, Astrophysical Journal v.487,
  487, 782

\bibitem[{Demorest {et~al.}(2012)Demorest, Ford, \& Ransom}]{Demorest:2012vx}
Demorest, P., Ford, J., \& Ransom, S. 2012, in prep

\bibitem[{Eatough {et~al.}(2009)Eatough, Keane, \& Lyne}]{2009MNRAS.395..410E}
Eatough, R.~P., Keane, E.~F., \& Lyne, A.~G. 2009, arXiv:0901.3993v1,
  astro-ph.GA

\bibitem[{Gnedin {et~al.}(2010)Gnedin, Brown, \& Geller}]{Gnedin:2010vr}
Gnedin, O.~Y., Brown, W.~R., \& Geller, M.~J. 2010, The Astrophysical Journal
  Letters

\bibitem[{Gruenspecht(2010)}]{Gruenspecht:2010tr}
Gruenspecht, H. 2010, Center for Strategic and International Studies

\bibitem[{Hollis {et~al.}(2004)Hollis, Jewell, Lovas, \&
  Remijan}]{Hollis:2004p2650}
Hollis, J.~M., Jewell, P.~R., Lovas, F.~J., \& Remijan, A. 2004, The
  Astrophysical Journal, 613, L45

\bibitem[{Iglesias-Groth(2011)}]{2011MNRAS.411.1857I}
Iglesias-Groth, S. 2011, Monthly Notices of the Royal Astronomical Society,
  411, 1857

\bibitem[{Kardashev(1964)}]{1964SvA.....8..217K}
Kardashev, N.~S. 1964, Soviet Astronomy, 8, 217

\bibitem[{Kasting {et~al.}(1993)Kasting, Whitmire, \&
  Reynolds}]{1993Icar..101..108K}
Kasting, J.~F., Whitmire, D.~P., \& Reynolds, R.~T. 1993, Icarus, 101, 108

\bibitem[{Korpela {et~al.}(2002)Korpela, Werthimer, \&
  Anderson}]{Korpela:2002p3225}
Korpela, E., Werthimer, D., \& Anderson, D. 2002, Computing in Science and
  Engineering

\bibitem[{Kuan {et~al.}(2003)Kuan, Charnley, Huang, Tseng, \&
  Kisiel}]{Kuan:2003eq}
Kuan, Y.~J., Charnley, S.~B., Huang, H.~C., Tseng, W.~L., \& Kisiel, Z. 2003,
  The Astrophysical Journal, 593, 848

\bibitem[{Leigh(1998)}]{Leigh:1998p12013}
Leigh, D. 1998, Harvard University PhD Thesis

\bibitem[{Lovas {et~al.}(2006)Lovas, Hollis, Remijan, \&
  Jewell}]{Lovas:2006p2493}
Lovas, F.~J., Hollis, J.~M., Remijan, A.~J., \& Jewell, P.~R. 2006, The
  Astrophysical Journal, 645, L137

\bibitem[{Marcy \& Howard(2011)}]{2011IAUS..276....3M}
Marcy, G.~W., \& Howard, A.~W. 2011, The Astrophysics of Planetary Systems:
  Formation, 276, 3

\bibitem[{Mehringer {et~al.}(1997)Mehringer, Snyder, Miao, \&
  Lovas}]{Mehringer:1997p2145}
Mehringer, D.~M., Snyder, L.~E., Miao, Y., \& Lovas, F.~J. 1997, Astrophysical
  Journal Letters v.480, 480, L71

\bibitem[{Morabito(2009)}]{2009RaSc...44.6004M}
Morabito, D.~D. 2009, Radio Science, 44, 6004

\bibitem[{Morrison {et~al.}(1977)Morrison, Billingham, \&
  Wolfe}]{Morrison:1977p182}
Morrison, P., Billingham, J., \& Wolfe, J. 1977, {The Search for
  Extraterrestrial Intelligence}

\bibitem[{Narayan(1992)}]{1992RSPTA.341..151N}
Narayan, R. 1992, Philosophical Transactions: Physical Sciences and
  Engineering, 341, 151

\bibitem[{Perillat(2012)}]{Perillat:2012ve}
Perillat, P. 2012

\bibitem[{Rickett(1990)}]{1990ARA&A..28..561R}
Rickett, B.~J. 1990, IN: Annual review of astronomy and astrophysics. Vol. 28
  (A91-28201 10-90). Palo Alto, 28, 561

\bibitem[{Robin {et~al.}(2003)Robin, Reyl{\'e}, Derri{\`e}re, \&
  Picaud}]{2003A&A...409..523R}
Robin, A.~C., Reyl{\'e}, C., Derri{\`e}re, S., \& Picaud, S. 2003, Astronomy
  and Astrophysics, 409, 523

\bibitem[{Schopf {et~al.}(2002)Schopf, Kudryavtsev, Agresti, Wdowiak, \&
  Czaja}]{Schopf:2002p2663}
Schopf, J.~W., Kudryavtsev, A.~B., Agresti, D.~G., Wdowiak, T.~J., \& Czaja,
  A.~D. 2002, Nature, 416, 73

\bibitem[{Siemion {et~al.}(2011)Siemion, Cobb, Chen, Cordes, Filiba, Foster,
  Fries, Howard, von Korff, Korpela, Lebofsky, McMahon, Parsons, Spitler,
  Wagner, \& Werthimer}]{2011arXiv1109.1136S}
Siemion, A. P.~V., {et~al.} 2011, arXiv.org, astro-ph.IM, 1109.1136

\bibitem[{Siemion {et~al.}(2012)Siemion, Bower, Foster, McMahon, Wagner,
  Werthimer, Backer, Cordes, \& van Leeuwen}]{2012ApJ...744..109S}
---. 2012, The Astrophysical Journal, 744, 109

\bibitem[{Taylor(1974)}]{1974A&AS...15..367T}
Taylor, J.~H. 1974, Astronomy and Astrophysics Supplement, 15, 367

\bibitem[{Woo(1978)}]{1978ApJ...219..727W}
Woo, R. 1978, Astrophysical Journal, 219, 727

\bibitem[{Woo(2007)}]{2007SpWea...509004W}
---. 2007, Space Weather, 5, 09004

\bibitem[{Woo \& Armstrong(1979)}]{1979JGR....84.7288W}
Woo, R., \& Armstrong, J.~W. 1979, Journal of Geophysical Research, 84, 7288

\end{thebibliography}
\appendix

\renewcommand{\thefootnote}{\alph{footnote}}

\begin{landscape}

\footnotesize
Table 3:: \Kepler Field Targeted Observations 02/2011$-$04/2011
\centering
\vspace{0.1in}
\begin{longtable}{|l|l l|l l|l l|l l|l l|l l|l l|}
\hline\hline
\textbf{Objects\footnote[1]{Objects in {\em italic} text were observed serendipitously with a primary target, see text.  KOI 326, one of our original targets, has been omitted from the table as its single identified planet candidate was determined to be a false positive.}} & \multicolumn{2}{|c|}{\textbf{Band 0}} & \multicolumn{2}{c|}{\textbf{Band 2}\footnote[2]{We exclude Band 1 here, as the band 1200-1330 MHz was not searched due to the presence of a bandpass filter used to mitigate heavy aircraft radar interference contaminating this region.}} & \multicolumn{2}{c|}{\textbf{Band 3}} & \multicolumn{2}{c|}{\textbf{Band 4}} & \multicolumn{2}{c|}{\textbf{Band 5}} & \multicolumn{2}{c|}{\textbf{Band 6}} & \multicolumn{2}{c|}{\textbf{Band 7}} \\
& \multicolumn{2}{|c|}{\textbf{1.1 -- 1.2 GHz}} & \multicolumn{2}{|c|}{\textbf{1.33 -- 1.4 GHz}}  &\multicolumn{2}{|c|}{\textbf{1.4 -- 1.5 GHz}}  & \multicolumn{2}{|c|}{\textbf{1.5 -- 1.6 GHz}}  & \multicolumn{2}{|c|}{\textbf{1.6 -- 1.7 GHz}}  & \multicolumn{2}{|c|}{\textbf{1.7 -- 1.8 GHz}}  &\multicolumn{2}{|c|}{\textbf{1.8 -- 1.9 GHz}} \\
 & $T_{\rm{total}}$ & \multicolumn{1}{c|}{$S_{\rm{peak}}$\footnote[3]{Peak sensitivity quoted for 0.75 Hz channelization}} & $T_{\rm{total}}$ & \multicolumn{1}{c|}{$S_{\rm{peak}}$} & $T_{\rm{total}}$ & \multicolumn{1}{c|}{$S_{\rm{peak}}$} & $T_{\rm{total}}$ & \multicolumn{1}{c|}{$S_{\rm{peak}}$} & $T_{\rm{total}}$ & \multicolumn{1}{c|}{$S_{\rm.{peak}}$} & $T_{\rm{total}}$ & \multicolumn{1}{c|}{$S_{\rm{peak}}$} & $T_{\rm{total}}$ & \multicolumn{1}{c|}{$S_{\rm{peak}}$} \\[0.02in]
 & \multicolumn{1}{c}{(s)} &($\frac{\displaystyle\rm{erg}}{\displaystyle\rm{s}\cdot \rm{cm}^{2}}$)\footnote[4]{\ / $10^{-23}$} & \multicolumn{1}{c}{(s)} & ($\frac{\displaystyle\rm{erg}}{\displaystyle\rm{s}\cdot \rm{cm}^{2}}$) &  \multicolumn{1}{c}{(s)} & ($\frac{\displaystyle\rm{erg}}{\displaystyle\rm{s}\cdot \rm{cm}^{2}}$) &  \multicolumn{1}{c}{(s)} & ($\frac{\displaystyle\rm{erg}}{\displaystyle\rm{s}\cdot \rm{cm}^{2}}$) &  \multicolumn{1}{c}{(s)} &($\frac{\displaystyle\rm{erg}}{\displaystyle\rm{s}\cdot \rm{cm}^{2}}$) &  \multicolumn{1}{c}{(s)} & ($\frac{\displaystyle\rm{erg}}{\displaystyle\rm{s}\cdot \rm{cm}^{2}}$) &  \multicolumn{1}{c}{(s)} & ($\frac{\displaystyle\rm{erg}}{\displaystyle\rm{s}\cdot \rm{cm}^{2}}$) \\[0.08in]
\hline
\textbf{Kepler-10 b} & \multirow{2}{*}{296} & \multirow{2}{*}{1.78} & \multirow{2}{*}{301} & \multirow{2}{*}{1.76} & \multirow{2}{*}{303} & \multirow{2}{*}{1.76} & \multirow{2}{*}{305} & \multirow{2}{*}{1.75} & \multirow{2}{*}{307} & \multirow{2}{*}{1.75} & \multirow{2}{*}{309} & \multirow{2}{*}{1.74} & \multirow{2}{*}{311} & \multirow{2}{*}{1.73} \\
\textbf{Kepler-10 c} & &  &  &  &  &  &  &  &  &  &  &  &  &  \\
\hline
\textbf{Kepler-11 b} & \multirow{6}{*}{295} & \multirow{6}{*}{1.78} & \multirow{6}{*}{299} & \multirow{6}{*}{1.77} & \multirow{6}{*}{301} & \multirow{6}{*}{1.76} & \multirow{6}{*}{303} & \multirow{6}{*}{1.75} & \multirow{6}{*}{305} & \multirow{6}{*}{1.75} & \multirow{6}{*}{307} & \multirow{6}{*}{1.74} & \multirow{6}{*}{309} & \multirow{6}{*}{1.74} \\
\textbf{Kepler-11 c} & &  &  &  &  &  &  &  &  &  &  &  &  &  \\
\textbf{Kepler-11 d} & &  &  &  &  &  &  &  &  &  &  &  &  &  \\
\textbf{Kepler-11 e} & &  &  &  &  &  &  &  &  &  &  &  &  &  \\
\textbf{Kepler-11 f} & &  &  &  &  &  &  &  &  &  &  &  &  &  \\
\textbf{Kepler-11 g} & &  &  &  &  &  &  &  &  &  &  &  &  &  \\
\hline
\textbf{Kepler-20 b} & \multirow{5}{*}{295} & \multirow{5}{*}{1.78} & \multirow{5}{*}{299} & \multirow{5}{*}{1.77} & \multirow{5}{*}{301} & \multirow{5}{*}{1.76} & \multirow{5}{*}{303} & \multirow{5}{*}{1.76} & \multirow{5}{*}{305} & \multirow{5}{*}{1.75} & \multirow{5}{*}{308} & \multirow{5}{*}{1.75} & \multirow{5}{*}{309} & \multirow{5}{*}{1.74} \\
\textbf{Kepler-20 c} & &  &  &  &  &  &  &  &  &  &  &  &  &  \\
\textbf{Kepler-20 d} & &  &  &  &  &  &  &  &  &  &  &  &  &  \\
\textbf{Kepler-20 e} & &  &  &  &  &  &  &  &  &  &  &  &  &  \\
\textbf{Kepler-20 f} & &  &  &  &  &  &  &  &  &  &  &  &  &  \\
\hline
\textbf{Kepler-22 b} & \multirow{1}{*}{295} & \multirow{1}{*}{1.78} & \multirow{1}{*}{300} & \multirow{1}{*}{1.77} & \multirow{1}{*}{302} & \multirow{1}{*}{1.76} & \multirow{1}{*}{304} & \multirow{1}{*}{1.76} & \multirow{1}{*}{306} & \multirow{1}{*}{1.75} & \multirow{1}{*}{308} & \multirow{1}{*}{1.75} & \multirow{1}{*}{ --} & \multirow{1}{*}{ -- } \\
\hline
\textbf{Kepler-30 b} & \multirow{3}{*}{295} & \multirow{3}{*}{1.78} & \multirow{3}{*}{300} & \multirow{3}{*}{1.77} & \multirow{3}{*}{302} & \multirow{3}{*}{1.76} & \multirow{3}{*}{304} & \multirow{3}{*}{1.75} & \multirow{3}{*}{306} & \multirow{3}{*}{1.75} & \multirow{3}{*}{308} & \multirow{3}{*}{1.74} & \multirow{3}{*}{310} & \multirow{3}{*}{1.74} \\
\textbf{Kepler-30 c} & &  &  &  &  &  &  &  &  &  &  &  &  &  \\
\textbf{Kepler-30 d} & &  &  &  &  &  &  &  &  &  &  &  &  &  \\
\hline
\textbf{Kepler-31 b} & \multirow{4}{*}{ --} & \multirow{4}{*}{ -- } & \multirow{4}{*}{300} & \multirow{4}{*}{1.77} & \multirow{4}{*}{ --} & \multirow{4}{*}{ -- } & \multirow{4}{*}{ --} & \multirow{4}{*}{ -- } & \multirow{4}{*}{ --} & \multirow{4}{*}{ -- } & \multirow{4}{*}{ --} & \multirow{4}{*}{ -- } & \multirow{4}{*}{ --} & \multirow{4}{*}{ -- } \\
\textbf{Kepler-31 c} & &  &  &  &  &  &  &  &  &  &  &  &  &  \\
\textbf{KOI 935.03} & &  &  &  &  &  &  &  &  &  &  &  &  &  \\
\textbf{KOI 935.04} & &  &  &  &  &  &  &  &  &  &  &  &  &  \\
\hline
\pagebreak
\hline
\textbf{Kepler-32 b} & \multirow{4}{*}{ --} & \multirow{4}{*}{ -- } & \multirow{4}{*}{ --} & \multirow{4}{*}{ -- } & \multirow{4}{*}{ --} & \multirow{4}{*}{ -- } & \multirow{4}{*}{ --} & \multirow{4}{*}{ -- } & \multirow{4}{*}{ --} & \multirow{4}{*}{ -- } & \multirow{4}{*}{302} & \multirow{4}{*}{1.75} & \multirow{4}{*}{305} & \multirow{4}{*}{1.74} \\
\textbf{Kepler-32 c} & &  &  &  &  &  &  &  &  &  &  &  &  &  \\
\textbf{KOI 952.03} & &  &  &  &  &  &  &  &  &  &  &  &  &  \\
\textbf{KOI 952.04} & &  &  &  &  &  &  &  &  &  &  &  &  &  \\
\hline
\textbf{KOI 51.01} & \multirow{1}{*}{293} & \multirow{1}{*}{1.78} & \multirow{1}{*}{297} & \multirow{1}{*}{1.77} & \multirow{1}{*}{300} & \multirow{1}{*}{1.77} & \multirow{1}{*}{302} & \multirow{1}{*}{1.76} & \multirow{1}{*}{304} & \multirow{1}{*}{1.75} & \multirow{1}{*}{306} & \multirow{1}{*}{1.75} & \multirow{1}{*}{308} & \multirow{1}{*}{1.74} \\
\hline
\textbf{KOI 111.01} & \multirow{3}{*}{ --} & \multirow{3}{*}{ -- } & \multirow{3}{*}{ --} & \multirow{3}{*}{ -- } & \multirow{3}{*}{301} & \multirow{3}{*}{1.76} & \multirow{3}{*}{303} & \multirow{3}{*}{1.76} & \multirow{3}{*}{305} & \multirow{3}{*}{1.75} & \multirow{3}{*}{307} & \multirow{3}{*}{1.75} & \multirow{3}{*}{309} & \multirow{3}{*}{1.74} \\
\textbf{KOI 111.02} & &  &  &  &  &  &  &  &  &  &  &  &  &  \\
\textbf{KOI 111.03} & &  &  &  &  &  &  &  &  &  &  &  &  &  \\
\hline
\textbf{KOI 113.01} & \multirow{1}{*}{293} & \multirow{1}{*}{1.78} & \multirow{1}{*}{298} & \multirow{1}{*}{1.77} & \multirow{1}{*}{300} & \multirow{1}{*}{1.77} & \multirow{1}{*}{302} & \multirow{1}{*}{1.76} & \multirow{1}{*}{304} & \multirow{1}{*}{1.75} & \multirow{1}{*}{306} & \multirow{1}{*}{1.75} & \multirow{1}{*}{308} & \multirow{1}{*}{1.74} \\
\hline
\textbf{KOI 174.01} & \multirow{1}{*}{293} & \multirow{1}{*}{1.79} & \multirow{1}{*}{298} & \multirow{1}{*}{1.77} & \multirow{1}{*}{300} & \multirow{1}{*}{1.77} & \multirow{1}{*}{302} & \multirow{1}{*}{1.76} & \multirow{1}{*}{304} & \multirow{1}{*}{1.76} & \multirow{1}{*}{306} & \multirow{1}{*}{1.75} & \multirow{1}{*}{308} & \multirow{1}{*}{1.75} \\
\hline
\textbf{KOI 211.01} & \multirow{1}{*}{295} & \multirow{1}{*}{1.78} & \multirow{1}{*}{299} & \multirow{1}{*}{1.77} & \multirow{1}{*}{ --} & \multirow{1}{*}{ -- } & \multirow{1}{*}{ --} & \multirow{1}{*}{ -- } & \multirow{1}{*}{ --} & \multirow{1}{*}{ -- } & \multirow{1}{*}{ --} & \multirow{1}{*}{ -- } & \multirow{1}{*}{306} & \multirow{1}{*}{1.74} \\
\hline
\textbf{KOI 260.01} & \multirow{2}{*}{295} & \multirow{2}{*}{1.78} & \multirow{2}{*}{299} & \multirow{2}{*}{1.77} & \multirow{2}{*}{301} & \multirow{2}{*}{1.76} & \multirow{2}{*}{303} & \multirow{2}{*}{1.76} & \multirow{2}{*}{306} & \multirow{2}{*}{1.75} & \multirow{2}{*}{308} & \multirow{2}{*}{1.75} & \multirow{2}{*}{310} & \multirow{2}{*}{1.74} \\
\textbf{KOI 260.02} & &  &  &  &  &  &  &  &  &  &  &  &  &  \\
\hline
\textbf{KOI 314.01} & \multirow{3}{*}{295} & \multirow{3}{*}{1.78} & \multirow{3}{*}{299} & \multirow{3}{*}{1.77} & \multirow{3}{*}{301} & \multirow{3}{*}{1.76} & \multirow{3}{*}{303} & \multirow{3}{*}{1.76} & \multirow{3}{*}{306} & \multirow{3}{*}{1.75} & \multirow{3}{*}{308} & \multirow{3}{*}{1.75} & \multirow{3}{*}{309} & \multirow{3}{*}{1.74} \\
\textbf{KOI 314.02} & &  &  &  &  &  &  &  &  &  &  &  &  &  \\
\textbf{KOI 314.03} & &  &  &  &  &  &  &  &  &  &  &  &  &  \\
\hline
\textbf{KOI 351.01} & \multirow{3}{*}{295} & \multirow{3}{*}{1.78} & \multirow{3}{*}{299} & \multirow{3}{*}{1.77} & \multirow{3}{*}{301} & \multirow{3}{*}{1.76} & \multirow{3}{*}{303} & \multirow{3}{*}{1.76} & \multirow{3}{*}{305} & \multirow{3}{*}{1.75} & \multirow{3}{*}{307} & \multirow{3}{*}{1.74} & \multirow{3}{*}{309} & \multirow{3}{*}{1.74} \\
\textbf{KOI 351.02} & &  &  &  &  &  &  &  &  &  &  &  &  &  \\
\textbf{KOI 351.03} & &  &  &  &  &  &  &  &  &  &  &  &  &  \\
\hline
\textbf{KOI 365.01} & \multirow{1}{*}{ --} & \multirow{1}{*}{ -- } & \multirow{1}{*}{297} & \multirow{1}{*}{1.78} & \multirow{1}{*}{ --} & \multirow{1}{*}{ -- } & \multirow{1}{*}{ --} & \multirow{1}{*}{ -- } & \multirow{1}{*}{ --} & \multirow{1}{*}{ -- } & \multirow{1}{*}{304} & \multirow{1}{*}{1.75} & \multirow{1}{*}{306} & \multirow{1}{*}{1.75} \\
\hline
\textbf{KOI 372.01} & \multirow{1}{*}{295} & \multirow{1}{*}{1.78} & \multirow{1}{*}{300} & \multirow{1}{*}{1.77} & \multirow{1}{*}{302} & \multirow{1}{*}{1.76} & \multirow{1}{*}{304} & \multirow{1}{*}{1.75} & \multirow{1}{*}{306} & \multirow{1}{*}{1.75} & \multirow{1}{*}{308} & \multirow{1}{*}{1.74} & \multirow{1}{*}{310} & \multirow{1}{*}{1.74} \\
\hline
\textbf{KOI 374.01} & \multirow{1}{*}{ --} & \multirow{1}{*}{ -- } & \multirow{1}{*}{ --} & \multirow{1}{*}{ -- } & \multirow{1}{*}{ --} & \multirow{1}{*}{ -- } & \multirow{1}{*}{ --} & \multirow{1}{*}{ -- } & \multirow{1}{*}{302} & \multirow{1}{*}{1.75} & \multirow{1}{*}{304} & \multirow{1}{*}{1.75} & \multirow{1}{*}{306} & \multirow{1}{*}{1.74} \\
\hline
\textbf{KOI 375.01} & \multirow{1}{*}{ --} & \multirow{1}{*}{ -- } & \multirow{1}{*}{ --} & \multirow{1}{*}{ -- } & \multirow{1}{*}{ --} & \multirow{1}{*}{ -- } & \multirow{1}{*}{ --} & \multirow{1}{*}{ -- } & \multirow{1}{*}{ --} & \multirow{1}{*}{ -- } & \multirow{1}{*}{304} & \multirow{1}{*}{1.75} & \multirow{1}{*}{306} & \multirow{1}{*}{1.74} \\
\hline
\textbf{KOI 386.01} & \multirow{2}{*}{296} & \multirow{2}{*}{1.78} & \multirow{2}{*}{301} & \multirow{2}{*}{1.77} & \multirow{2}{*}{303} & \multirow{2}{*}{1.76} & \multirow{2}{*}{305} & \multirow{2}{*}{1.75} & \multirow{2}{*}{307} & \multirow{2}{*}{1.75} & \multirow{2}{*}{309} & \multirow{2}{*}{1.74} & \multirow{2}{*}{311} & \multirow{2}{*}{1.74} \\
\textbf{KOI 386.02} & &  &  &  &  &  &  &  &  &  &  &  &  &  \\
\hline
\textbf{KOI 401.01} & \multirow{2}{*}{296} & \multirow{2}{*}{1.78} & \multirow{2}{*}{301} & \multirow{2}{*}{1.76} & \multirow{2}{*}{303} & \multirow{2}{*}{1.76} & \multirow{2}{*}{289} & \multirow{2}{*}{1.79} & \multirow{2}{*}{307} & \multirow{2}{*}{1.75} & \multirow{2}{*}{309} & \multirow{2}{*}{1.74} & \multirow{2}{*}{311} & \multirow{2}{*}{1.73} \\
\textbf{KOI 401.02} & &  &  &  &  &  &  &  &  &  &  &  &  &  \\
\hline
\textbf{KOI 416.01} & \multirow{2}{*}{297} & \multirow{2}{*}{1.78} & \multirow{2}{*}{301} & \multirow{2}{*}{1.77} & \multirow{2}{*}{603} & \multirow{2}{*}{1.76} & \multirow{2}{*}{607} & \multirow{2}{*}{1.75} & \multirow{2}{*}{612} & \multirow{2}{*}{1.75} & \multirow{2}{*}{616} & \multirow{2}{*}{1.74} & \multirow{2}{*}{620} & \multirow{2}{*}{1.73} \\
\textbf{KOI 416.02} & &  &  &  &  &  &  &  &  &  &  &  &  &  \\
\hline
\textbf{KOI 422.01} & \multirow{1}{*}{ --} & \multirow{1}{*}{ -- } & \multirow{1}{*}{ --} & \multirow{1}{*}{ -- } & \multirow{1}{*}{ --} & \multirow{1}{*}{ -- } & \multirow{1}{*}{ --} & \multirow{1}{*}{ -- } & \multirow{1}{*}{305} & \multirow{1}{*}{1.75} & \multirow{1}{*}{307} & \multirow{1}{*}{1.75} & \multirow{1}{*}{310} & \multirow{1}{*}{1.74} \\
\hline
\textbf{KOI 433.01} & \multirow{2}{*}{294} & \multirow{2}{*}{1.78} & \multirow{2}{*}{299} & \multirow{2}{*}{1.77} & \multirow{2}{*}{301} & \multirow{2}{*}{1.76} & \multirow{2}{*}{303} & \multirow{2}{*}{1.76} & \multirow{2}{*}{305} & \multirow{2}{*}{1.75} & \multirow{2}{*}{307} & \multirow{2}{*}{1.75} & \multirow{2}{*}{309} & \multirow{2}{*}{1.74} \\
\textbf{KOI 433.02} & &  &  &  &  &  &  &  &  &  &  &  &  &  \\
\hline
\textbf{KOI 448.01} & \multirow{2}{*}{295} & \multirow{2}{*}{1.78} & \multirow{2}{*}{299} & \multirow{2}{*}{1.77} & \multirow{2}{*}{301} & \multirow{2}{*}{1.77} & \multirow{2}{*}{303} & \multirow{2}{*}{1.76} & \multirow{2}{*}{305} & \multirow{2}{*}{1.75} & \multirow{2}{*}{307} & \multirow{2}{*}{1.75} & \multirow{2}{*}{309} & \multirow{2}{*}{1.74} \\
\textbf{KOI 448.02} & &  &  &  &  &  &  &  &  &  &  &  &  &  \\
\hline
\textbf{KOI 465.01} & \multirow{1}{*}{ --} & \multirow{1}{*}{ -- } & \multirow{1}{*}{299} & \multirow{1}{*}{1.77} & \multirow{1}{*}{ --} & \multirow{1}{*}{ -- } & \multirow{1}{*}{ --} & \multirow{1}{*}{ -- } & \multirow{1}{*}{ --} & \multirow{1}{*}{ -- } & \multirow{1}{*}{ --} & \multirow{1}{*}{ -- } & \multirow{1}{*}{ --} & \multirow{1}{*}{ -- } \\
\hline
\pagebreak
\hline
\textbf{KOI 500.01} & \multirow{5}{*}{295} & \multirow{5}{*}{1.78} & \multirow{5}{*}{300} & \multirow{5}{*}{1.77} & \multirow{5}{*}{302} & \multirow{5}{*}{1.76} & \multirow{5}{*}{304} & \multirow{5}{*}{1.75} & \multirow{5}{*}{306} & \multirow{5}{*}{1.75} & \multirow{5}{*}{308} & \multirow{5}{*}{1.74} & \multirow{5}{*}{310} & \multirow{5}{*}{1.74} \\
\textbf{KOI 500.02} & &  &  &  &  &  &  &  &  &  &  &  &  &  \\
\textbf{KOI 500.03} & &  &  &  &  &  &  &  &  &  &  &  &  &  \\
\textbf{KOI 500.04} & &  &  &  &  &  &  &  &  &  &  &  &  &  \\
\textbf{KOI 500.05} & &  &  &  &  &  &  &  &  &  &  &  &  &  \\
\hline
\textbf{KOI 536.01} & \multirow{1}{*}{294} & \multirow{1}{*}{1.78} & \multirow{1}{*}{299} & \multirow{1}{*}{1.77} & \multirow{1}{*}{ --} & \multirow{1}{*}{ -- } & \multirow{1}{*}{ --} & \multirow{1}{*}{ -- } & \multirow{1}{*}{ --} & \multirow{1}{*}{ -- } & \multirow{1}{*}{ --} & \multirow{1}{*}{ -- } & \multirow{1}{*}{ --} & \multirow{1}{*}{ -- } \\
\hline
\textbf{\em{KOI 542.01}} & \multirow{2}{*}{ --} & \multirow{2}{*}{ -- } & \multirow{2}{*}{300} & \multirow{2}{*}{1.76} & \multirow{2}{*}{ --} & \multirow{2}{*}{ -- } & \multirow{2}{*}{303} & \multirow{2}{*}{1.75} & \multirow{2}{*}{305} & \multirow{2}{*}{1.75} & \multirow{2}{*}{307} & \multirow{2}{*}{1.74} & \multirow{2}{*}{309} & \multirow{2}{*}{1.74} \\
\textbf{\em{KOI 542.02}} & &  &  &  &  &  &  &  &  &  &  &  &  &  \\
\hline
\textbf{KOI 555.01} & \multirow{2}{*}{295} & \multirow{2}{*}{1.78} & \multirow{2}{*}{299} & \multirow{2}{*}{1.77} & \multirow{2}{*}{301} & \multirow{2}{*}{1.76} & \multirow{2}{*}{294} & \multirow{2}{*}{1.78} & \multirow{2}{*}{306} & \multirow{2}{*}{1.75} & \multirow{2}{*}{308} & \multirow{2}{*}{1.75} & \multirow{2}{*}{ --} & \multirow{2}{*}{ -- } \\
\textbf{KOI 555.02} & &  &  &  &  &  &  &  &  &  &  &  &  &  \\
\hline
\textbf{KOI 564.01} & \multirow{2}{*}{ --} & \multirow{2}{*}{ -- } & \multirow{2}{*}{ --} & \multirow{2}{*}{ -- } & \multirow{2}{*}{300} & \multirow{2}{*}{1.76} & \multirow{2}{*}{287} & \multirow{2}{*}{1.77} & \multirow{2}{*}{304} & \multirow{2}{*}{1.75} & \multirow{2}{*}{306} & \multirow{2}{*}{1.75} & \multirow{2}{*}{308} & \multirow{2}{*}{1.74} \\
\textbf{KOI 564.02} & &  &  &  &  &  &  &  &  &  &  &  &  &  \\
\hline
\textbf{KOI 590.01} & \multirow{2}{*}{295} & \multirow{2}{*}{1.78} & \multirow{2}{*}{299} & \multirow{2}{*}{1.77} & \multirow{2}{*}{ --} & \multirow{2}{*}{ -- } & \multirow{2}{*}{ --} & \multirow{2}{*}{ -- } & \multirow{2}{*}{ --} & \multirow{2}{*}{ -- } & \multirow{2}{*}{308} & \multirow{2}{*}{1.75} & \multirow{2}{*}{310} & \multirow{2}{*}{1.74} \\
\textbf{KOI 590.02} & &  &  &  &  &  &  &  &  &  &  &  &  &  \\
\hline
\textbf{\em{KOI 618.01}} & \multirow{1}{*}{ --} & \multirow{1}{*}{ -- } & \multirow{1}{*}{ --} & \multirow{1}{*}{ -- } & \multirow{1}{*}{ --} & \multirow{1}{*}{ -- } & \multirow{1}{*}{ --} & \multirow{1}{*}{ -- } & \multirow{1}{*}{ --} & \multirow{1}{*}{ -- } & \multirow{1}{*}{303} & \multirow{1}{*}{1.75} & \multirow{1}{*}{305} & \multirow{1}{*}{1.75} \\
\hline
\textbf{KOI 622.01} & \multirow{1}{*}{ --} & \multirow{1}{*}{ -- } & \multirow{1}{*}{ --} & \multirow{1}{*}{ -- } & \multirow{1}{*}{ --} & \multirow{1}{*}{ -- } & \multirow{1}{*}{ --} & \multirow{1}{*}{ -- } & \multirow{1}{*}{ --} & \multirow{1}{*}{ -- } & \multirow{1}{*}{301} & \multirow{1}{*}{1.75} & \multirow{1}{*}{304} & \multirow{1}{*}{1.75} \\
\hline
\textbf{KOI 682.01} & \multirow{1}{*}{295} & \multirow{1}{*}{1.78} & \multirow{1}{*}{599} & \multirow{1}{*}{1.77} & \multirow{1}{*}{302} & \multirow{1}{*}{1.76} & \multirow{1}{*}{304} & \multirow{1}{*}{1.75} & \multirow{1}{*}{306} & \multirow{1}{*}{1.75} & \multirow{1}{*}{308} & \multirow{1}{*}{1.74} & \multirow{1}{*}{310} & \multirow{1}{*}{1.74} \\
\hline
\textbf{KOI 683.01} & \multirow{1}{*}{296} & \multirow{1}{*}{1.78} & \multirow{1}{*}{300} & \multirow{1}{*}{1.77} & \multirow{1}{*}{302} & \multirow{1}{*}{1.76} & \multirow{1}{*}{304} & \multirow{1}{*}{1.75} & \multirow{1}{*}{306} & \multirow{1}{*}{1.75} & \multirow{1}{*}{308} & \multirow{1}{*}{1.74} & \multirow{1}{*}{310} & \multirow{1}{*}{1.74} \\
\hline
\textbf{\em{KOI 698.01}} & \multirow{1}{*}{ --} & \multirow{1}{*}{ -- } & \multirow{1}{*}{299} & \multirow{1}{*}{1.77} & \multirow{1}{*}{ --} & \multirow{1}{*}{ -- } & \multirow{1}{*}{ --} & \multirow{1}{*}{ -- } & \multirow{1}{*}{ --} & \multirow{1}{*}{ -- } & \multirow{1}{*}{ --} & \multirow{1}{*}{ -- } & \multirow{1}{*}{ --} & \multirow{1}{*}{ -- } \\
\hline
\textbf{KOI 701.01} & \multirow{3}{*}{ --} & \multirow{3}{*}{ -- } & \multirow{3}{*}{ --} & \multirow{3}{*}{ -- } & \multirow{3}{*}{ --} & \multirow{3}{*}{ -- } & \multirow{3}{*}{ --} & \multirow{3}{*}{ -- } & \multirow{3}{*}{301} & \multirow{3}{*}{1.75} & \multirow{3}{*}{303} & \multirow{3}{*}{1.75} & \multirow{3}{*}{305} & \multirow{3}{*}{1.74} \\
\textbf{KOI 701.02} & &  &  &  &  &  &  &  &  &  &  &  &  &  \\
\textbf{KOI 701.03} & &  &  &  &  &  &  &  &  &  &  &  &  &  \\
\hline
\textbf{KOI 711.01} & \multirow{3}{*}{ --} & \multirow{3}{*}{ -- } & \multirow{3}{*}{ --} & \multirow{3}{*}{ -- } & \multirow{3}{*}{302} & \multirow{3}{*}{1.76} & \multirow{3}{*}{304} & \multirow{3}{*}{1.76} & \multirow{3}{*}{306} & \multirow{3}{*}{1.75} & \multirow{3}{*}{308} & \multirow{3}{*}{1.75} & \multirow{3}{*}{310} & \multirow{3}{*}{1.74} \\
\textbf{KOI 711.02} & &  &  &  &  &  &  &  &  &  &  &  &  &  \\
\textbf{KOI 711.03} & &  &  &  &  &  &  &  &  &  &  &  &  &  \\
\hline
\textbf{\em{KOI 741.01}} & \multirow{1}{*}{ --} & \multirow{1}{*}{ -- } & \multirow{1}{*}{ --} & \multirow{1}{*}{ -- } & \multirow{1}{*}{ --} & \multirow{1}{*}{ -- } & \multirow{1}{*}{ --} & \multirow{1}{*}{ -- } & \multirow{1}{*}{ --} & \multirow{1}{*}{ -- } & \multirow{1}{*}{303} & \multirow{1}{*}{1.75} & \multirow{1}{*}{305} & \multirow{1}{*}{1.75} \\
\hline
\textbf{KOI 812.01} & \multirow{4}{*}{296} & \multirow{4}{*}{1.78} & \multirow{4}{*}{301} & \multirow{4}{*}{1.76} & \multirow{4}{*}{303} & \multirow{4}{*}{1.76} & \multirow{4}{*}{295} & \multirow{4}{*}{1.78} & \multirow{4}{*}{307} & \multirow{4}{*}{1.75} & \multirow{4}{*}{309} & \multirow{4}{*}{1.74} & \multirow{4}{*}{311} & \multirow{4}{*}{1.73} \\
\textbf{KOI 812.02} & &  &  &  &  &  &  &  &  &  &  &  &  &  \\
\textbf{KOI 812.03} & &  &  &  &  &  &  &  &  &  &  &  &  &  \\
\textbf{KOI 812.04} & &  &  &  &  &  &  &  &  &  &  &  &  &  \\
\hline
\textbf{KOI 817.01} & \multirow{2}{*}{296} & \multirow{2}{*}{1.78} & \multirow{2}{*}{300} & \multirow{2}{*}{1.76} & \multirow{2}{*}{302} & \multirow{2}{*}{1.76} & \multirow{2}{*}{292} & \multirow{2}{*}{1.78} & \multirow{2}{*}{307} & \multirow{2}{*}{1.75} & \multirow{2}{*}{309} & \multirow{2}{*}{1.74} & \multirow{2}{*}{311} & \multirow{2}{*}{1.73} \\
\textbf{KOI 817.02} & &  &  &  &  &  &  &  &  &  &  &  &  &  \\
\hline
\textbf{\em{KOI 826.01}} & \multirow{1}{*}{296} & \multirow{1}{*}{1.78} & \multirow{1}{*}{300} & \multirow{1}{*}{1.77} & \multirow{1}{*}{302} & \multirow{1}{*}{1.76} & \multirow{1}{*}{295} & \multirow{1}{*}{1.76} & \multirow{1}{*}{306} & \multirow{1}{*}{1.75} & \multirow{1}{*}{308} & \multirow{1}{*}{1.74} & \multirow{1}{*}{ --} & \multirow{1}{*}{ -- } \\
\hline
\textbf{KOI 847.01} & \multirow{1}{*}{294} & \multirow{1}{*}{1.78} & \multirow{1}{*}{298} & \multirow{1}{*}{1.77} & \multirow{1}{*}{301} & \multirow{1}{*}{1.77} & \multirow{1}{*}{303} & \multirow{1}{*}{1.76} & \multirow{1}{*}{305} & \multirow{1}{*}{1.75} & \multirow{1}{*}{307} & \multirow{1}{*}{1.75} & \multirow{1}{*}{309} & \multirow{1}{*}{1.74} \\
\hline
\textbf{KOI 854.01} & \multirow{1}{*}{295} & \multirow{1}{*}{1.78} & \multirow{1}{*}{299} & \multirow{1}{*}{1.77} & \multirow{1}{*}{301} & \multirow{1}{*}{1.76} & \multirow{1}{*}{304} & \multirow{1}{*}{1.76} & \multirow{1}{*}{306} & \multirow{1}{*}{1.75} & \multirow{1}{*}{308} & \multirow{1}{*}{1.75} & \multirow{1}{*}{310} & \multirow{1}{*}{1.74} \\
\hline
\textbf{KOI 882.01} & \multirow{1}{*}{294} & \multirow{1}{*}{1.78} & \multirow{1}{*}{299} & \multirow{1}{*}{1.77} & \multirow{1}{*}{301} & \multirow{1}{*}{1.76} & \multirow{1}{*}{303} & \multirow{1}{*}{1.76} & \multirow{1}{*}{305} & \multirow{1}{*}{1.75} & \multirow{1}{*}{307} & \multirow{1}{*}{1.75} & \multirow{1}{*}{309} & \multirow{1}{*}{1.74} \\
\hline
\textbf{\em{KOI 892.01}} & \multirow{1}{*}{295} & \multirow{1}{*}{1.78} & \multirow{1}{*}{299} & \multirow{1}{*}{1.77} & \multirow{1}{*}{301} & \multirow{1}{*}{1.76} & \multirow{1}{*}{303} & \multirow{1}{*}{1.76} & \multirow{1}{*}{306} & \multirow{1}{*}{1.75} & \multirow{1}{*}{308} & \multirow{1}{*}{1.75} & \multirow{1}{*}{309} & \multirow{1}{*}{1.74} \\
\hline
\textbf{KOI 902.01} & \multirow{1}{*}{ --} & \multirow{1}{*}{ -- } & \multirow{1}{*}{ --} & \multirow{1}{*}{ -- } & \multirow{1}{*}{299} & \multirow{1}{*}{1.76} & \multirow{1}{*}{301} & \multirow{1}{*}{1.76} & \multirow{1}{*}{303} & \multirow{1}{*}{1.75} & \multirow{1}{*}{305} & \multirow{1}{*}{1.75} & \multirow{1}{*}{308} & \multirow{1}{*}{1.74} \\
\hline
\textbf{KOI 947.01} & \multirow{1}{*}{ --} & \multirow{1}{*}{ -- } & \multirow{1}{*}{ --} & \multirow{1}{*}{ -- } & \multirow{1}{*}{ --} & \multirow{1}{*}{ -- } & \multirow{1}{*}{294} & \multirow{1}{*}{1.77} & \multirow{1}{*}{311} & \multirow{1}{*}{1.72} & \multirow{1}{*}{313} & \multirow{1}{*}{1.72} & \multirow{1}{*}{315} & \multirow{1}{*}{1.71} \\
\hline
\textbf{KOI 974.01} & \multirow{1}{*}{294} & \multirow{1}{*}{1.78} & \multirow{1}{*}{299} & \multirow{1}{*}{1.77} & \multirow{1}{*}{ --} & \multirow{1}{*}{ -- } & \multirow{1}{*}{ --} & \multirow{1}{*}{ -- } & \multirow{1}{*}{ --} & \multirow{1}{*}{ -- } & \multirow{1}{*}{304} & \multirow{1}{*}{1.75} & \multirow{1}{*}{306} & \multirow{1}{*}{1.75} \\
\hline
\textbf{\em{KOI 986.01}} & \multirow{2}{*}{296} & \multirow{2}{*}{1.77} & \multirow{2}{*}{301} & \multirow{2}{*}{1.76} & \multirow{2}{*}{303} & \multirow{2}{*}{1.76} & \multirow{2}{*}{299} & \multirow{2}{*}{1.75} & \multirow{2}{*}{307} & \multirow{2}{*}{1.74} & \multirow{2}{*}{309} & \multirow{2}{*}{1.74} & \multirow{2}{*}{311} & \multirow{2}{*}{1.73} \\
\textbf{\em{KOI 986.02}} & &  &  &  &  &  &  &  &  &  &  &  &  &  \\
\hline
\textbf{KOI 998.01} & \multirow{1}{*}{294} & \multirow{1}{*}{1.78} & \multirow{1}{*}{298} & \multirow{1}{*}{1.77} & \multirow{1}{*}{300} & \multirow{1}{*}{1.77} & \multirow{1}{*}{302} & \multirow{1}{*}{1.76} & \multirow{1}{*}{304} & \multirow{1}{*}{1.75} & \multirow{1}{*}{306} & \multirow{1}{*}{1.75} & \multirow{1}{*}{308} & \multirow{1}{*}{1.74} \\
\hline
\textbf{KOI 1010.01} & \multirow{1}{*}{294} & \multirow{1}{*}{1.78} & \multirow{1}{*}{298} & \multirow{1}{*}{1.77} & \multirow{1}{*}{300} & \multirow{1}{*}{1.77} & \multirow{1}{*}{302} & \multirow{1}{*}{1.76} & \multirow{1}{*}{304} & \multirow{1}{*}{1.75} & \multirow{1}{*}{306} & \multirow{1}{*}{1.75} & \multirow{1}{*}{308} & \multirow{1}{*}{1.74} \\
\hline
\textbf{KOI 1032.01} & \multirow{1}{*}{294} & \multirow{1}{*}{1.78} & \multirow{1}{*}{298} & \multirow{1}{*}{1.77} & \multirow{1}{*}{300} & \multirow{1}{*}{1.77} & \multirow{1}{*}{302} & \multirow{1}{*}{1.76} & \multirow{1}{*}{304} & \multirow{1}{*}{1.75} & \multirow{1}{*}{306} & \multirow{1}{*}{1.75} & \multirow{1}{*}{308} & \multirow{1}{*}{1.74} \\
\hline
\textbf{KOI 1099.01} & \multirow{1}{*}{296} & \multirow{1}{*}{1.78} & \multirow{1}{*}{301} & \multirow{1}{*}{1.76} & \multirow{1}{*}{303} & \multirow{1}{*}{1.76} & \multirow{1}{*}{305} & \multirow{1}{*}{1.75} & \multirow{1}{*}{307} & \multirow{1}{*}{1.75} & \multirow{1}{*}{309} & \multirow{1}{*}{1.74} & \multirow{1}{*}{311} & \multirow{1}{*}{1.73} \\
\hline
\textbf{KOI 1113.01} & \multirow{2}{*}{296} & \multirow{2}{*}{1.77} & \multirow{2}{*}{301} & \multirow{2}{*}{1.76} & \multirow{2}{*}{303} & \multirow{2}{*}{1.76} & \multirow{2}{*}{299} & \multirow{2}{*}{1.75} & \multirow{2}{*}{307} & \multirow{2}{*}{1.74} & \multirow{2}{*}{309} & \multirow{2}{*}{1.74} & \multirow{2}{*}{311} & \multirow{2}{*}{1.73} \\
\textbf{KOI 1113.02} & &  &  &  &  &  &  &  &  &  &  &  &  &  \\
\hline
\textbf{\em{KOI 1118.01}} & \multirow{1}{*}{296} & \multirow{1}{*}{1.78} & \multirow{1}{*}{301} & \multirow{1}{*}{1.76} & \multirow{1}{*}{303} & \multirow{1}{*}{1.76} & \multirow{1}{*}{305} & \multirow{1}{*}{1.75} & \multirow{1}{*}{307} & \multirow{1}{*}{1.75} & \multirow{1}{*}{309} & \multirow{1}{*}{1.74} & \multirow{1}{*}{311} & \multirow{1}{*}{1.73} \\
\hline
\textbf{KOI 1159.01} & \multirow{1}{*}{ --} & \multirow{1}{*}{ -- } & \multirow{1}{*}{ --} & \multirow{1}{*}{ -- } & \multirow{1}{*}{ --} & \multirow{1}{*}{ -- } & \multirow{1}{*}{ --} & \multirow{1}{*}{ -- } & \multirow{1}{*}{ --} & \multirow{1}{*}{ -- } & \multirow{1}{*}{303} & \multirow{1}{*}{1.75} & \multirow{1}{*}{305} & \multirow{1}{*}{1.75} \\
\hline
\textbf{KOI 1162.01} & \multirow{1}{*}{296} & \multirow{1}{*}{1.78} & \multirow{1}{*}{300} & \multirow{1}{*}{1.77} & \multirow{1}{*}{302} & \multirow{1}{*}{1.76} & \multirow{1}{*}{ --} & \multirow{1}{*}{ -- } & \multirow{1}{*}{ --} & \multirow{1}{*}{ -- } & \multirow{1}{*}{ --} & \multirow{1}{*}{ -- } & \multirow{1}{*}{ --} & \multirow{1}{*}{ -- } \\
\hline
\textbf{KOI 1168.01} & \multirow{1}{*}{ --} & \multirow{1}{*}{ -- } & \multirow{1}{*}{300} & \multirow{1}{*}{1.77} & \multirow{1}{*}{302} & \multirow{1}{*}{1.76} & \multirow{1}{*}{ --} & \multirow{1}{*}{ -- } & \multirow{1}{*}{ --} & \multirow{1}{*}{ -- } & \multirow{1}{*}{ --} & \multirow{1}{*}{ -- } & \multirow{1}{*}{ --} & \multirow{1}{*}{ -- } \\
\hline
\textbf{KOI 1192.01} & \multirow{1}{*}{294} & \multirow{1}{*}{1.78} & \multirow{1}{*}{298} & \multirow{1}{*}{1.77} & \multirow{1}{*}{ --} & \multirow{1}{*}{ -- } & \multirow{1}{*}{ --} & \multirow{1}{*}{ -- } & \multirow{1}{*}{ --} & \multirow{1}{*}{ -- } & \multirow{1}{*}{ --} & \multirow{1}{*}{ -- } & \multirow{1}{*}{ --} & \multirow{1}{*}{ -- } \\
\hline
\textbf{KOI 1199.01} & \multirow{1}{*}{295} & \multirow{1}{*}{1.78} & \multirow{1}{*}{299} & \multirow{1}{*}{1.77} & \multirow{1}{*}{301} & \multirow{1}{*}{1.77} & \multirow{1}{*}{303} & \multirow{1}{*}{1.76} & \multirow{1}{*}{305} & \multirow{1}{*}{1.75} & \multirow{1}{*}{307} & \multirow{1}{*}{1.75} & \multirow{1}{*}{309} & \multirow{1}{*}{1.74} \\
\hline
\textbf{\em{KOI 1203.01}}& \multirow{3}{*}{295} & \multirow{3}{*}{1.78} & \multirow{3}{*}{299} & \multirow{3}{*}{1.77} & \multirow{3}{*}{301} & \multirow{3}{*}{1.76} & \multirow{3}{*}{303} & \multirow{3}{*}{1.76} & \multirow{3}{*}{305} & \multirow{3}{*}{1.75} & \multirow{3}{*}{307} & \multirow{3}{*}{1.75} & \multirow{3}{*}{309} & \multirow{3}{*}{1.74} \\
\textbf{\em{KOI 1203.02}} & &  &  &  &  &  &  &  &  &  &  &  &  &  \\
\textbf{\em{KOI 1203.03}} & &  &  &  &  &  &  &  &  &  &  &  &  &  \\
\hline
\textbf{KOI 1208.01} & \multirow{1}{*}{295} & \multirow{1}{*}{1.78} & \multirow{1}{*}{299} & \multirow{1}{*}{1.77} & \multirow{1}{*}{301} & \multirow{1}{*}{1.76} & \multirow{1}{*}{303} & \multirow{1}{*}{1.76} & \multirow{1}{*}{305} & \multirow{1}{*}{1.75} & \multirow{1}{*}{307} & \multirow{1}{*}{1.75} & \multirow{1}{*}{309} & \multirow{1}{*}{1.74} \\
\hline
\textbf{\em{KOI 1210.01}} & \multirow{1}{*}{295} & \multirow{1}{*}{1.78} & \multirow{1}{*}{299} & \multirow{1}{*}{1.77} & \multirow{1}{*}{301} & \multirow{1}{*}{1.76} & \multirow{1}{*}{303} & \multirow{1}{*}{1.76} & \multirow{1}{*}{305} & \multirow{1}{*}{1.75} & \multirow{1}{*}{307} & \multirow{1}{*}{1.75} & \multirow{1}{*}{309} & \multirow{1}{*}{1.74} \\
\hline
\textbf{KOI 1226.01} & \multirow{1}{*}{295} & \multirow{1}{*}{1.78} & \multirow{1}{*}{299} & \multirow{1}{*}{1.77} & \multirow{1}{*}{301} & \multirow{1}{*}{1.77} & \multirow{1}{*}{303} & \multirow{1}{*}{1.76} & \multirow{1}{*}{305} & \multirow{1}{*}{1.75} & \multirow{1}{*}{307} & \multirow{1}{*}{1.75} & \multirow{1}{*}{309} & \multirow{1}{*}{1.74} \\
\hline
\textbf{KOI 1261.01} & \multirow{2}{*}{ --} & \multirow{2}{*}{ -- } & \multirow{2}{*}{ --} & \multirow{2}{*}{ -- } & \multirow{2}{*}{ --} & \multirow{2}{*}{ -- } & \multirow{2}{*}{ --} & \multirow{2}{*}{ -- } & \multirow{2}{*}{ --} & \multirow{2}{*}{ -- } & \multirow{2}{*}{307} & \multirow{2}{*}{1.75} & \multirow{2}{*}{309} & \multirow{2}{*}{1.74} \\
\textbf{KOI 1261.02} & &  &  &  &  &  &  &  &  &  &  &  &  &  \\
\hline
\textbf{KOI 1268.01} & \multirow{1}{*}{295} & \multirow{1}{*}{1.78} & \multirow{1}{*}{299} & \multirow{1}{*}{1.77} & \multirow{1}{*}{301} & \multirow{1}{*}{1.76} & \multirow{1}{*}{303} & \multirow{1}{*}{1.76} & \multirow{1}{*}{306} & \multirow{1}{*}{1.75} & \multirow{1}{*}{308} & \multirow{1}{*}{1.75} & \multirow{1}{*}{310} & \multirow{1}{*}{1.74} \\
\hline
\textbf{KOI 1302.01} & \multirow{1}{*}{292} & \multirow{1}{*}{1.79} & \multirow{1}{*}{296} & \multirow{1}{*}{1.78} & \multirow{1}{*}{298} & \multirow{1}{*}{1.77} & \multirow{1}{*}{300} & \multirow{1}{*}{1.76} & \multirow{1}{*}{303} & \multirow{1}{*}{1.76} & \multirow{1}{*}{305} & \multirow{1}{*}{1.75} & \multirow{1}{*}{ --} & \multirow{1}{*}{ -- } \\
\hline
\textbf{KOI 1328.01} & \multirow{1}{*}{296} & \multirow{1}{*}{1.78} & \multirow{1}{*}{300} & \multirow{1}{*}{1.76} & \multirow{1}{*}{302} & \multirow{1}{*}{1.76} & \multirow{1}{*}{304} & \multirow{1}{*}{1.75} & \multirow{1}{*}{306} & \multirow{1}{*}{1.75} & \multirow{1}{*}{309} & \multirow{1}{*}{1.74} & \multirow{1}{*}{310} & \multirow{1}{*}{1.73} \\
\hline
\textbf{KOI 1355.01} & \multirow{1}{*}{297} & \multirow{1}{*}{1.78} & \multirow{1}{*}{599} & \multirow{1}{*}{1.77} & \multirow{1}{*}{603} & \multirow{1}{*}{1.76} & \multirow{1}{*}{600} & \multirow{1}{*}{1.75} & \multirow{1}{*}{611} & \multirow{1}{*}{1.75} & \multirow{1}{*}{616} & \multirow{1}{*}{1.74} & \multirow{1}{*}{620} & \multirow{1}{*}{1.74} \\
\hline
\textbf{\em{KOI 1358.01}} & \multirow{3}{*}{294} & \multirow{3}{*}{1.78} & \multirow{3}{*}{299} & \multirow{3}{*}{1.77} & \multirow{3}{*}{301} & \multirow{3}{*}{1.76} & \multirow{3}{*}{303} & \multirow{3}{*}{1.76} & \multirow{3}{*}{305} & \multirow{3}{*}{1.75} & \multirow{3}{*}{307} & \multirow{3}{*}{1.75} & \multirow{3}{*}{309} & \multirow{3}{*}{1.74} \\
\textbf{\em{KOI 1358.02}} & &  &  &  &  &  &  &  &  &  &  &  &  &  \\
\textbf{\em{KOI 1358.03}} & &  &  &  &  &  &  &  &  &  &  &  &  &  \\
\hline
\textbf{KOI 1361.01} & \multirow{1}{*}{ --} & \multirow{1}{*}{ -- } & \multirow{1}{*}{ --} & \multirow{1}{*}{ -- } & \multirow{1}{*}{302} & \multirow{1}{*}{1.76} & \multirow{1}{*}{295} & \multirow{1}{*}{1.78} & \multirow{1}{*}{306} & \multirow{1}{*}{1.75} & \multirow{1}{*}{308} & \multirow{1}{*}{1.75} & \multirow{1}{*}{310} & \multirow{1}{*}{1.74} \\
\hline
\textbf{KOI 1372.01} & \multirow{1}{*}{296} & \multirow{1}{*}{1.78} & \multirow{1}{*}{300} & \multirow{1}{*}{1.76} & \multirow{1}{*}{302} & \multirow{1}{*}{1.76} & \multirow{1}{*}{304} & \multirow{1}{*}{1.75} & \multirow{1}{*}{306} & \multirow{1}{*}{1.75} & \multirow{1}{*}{308} & \multirow{1}{*}{1.74} & \multirow{1}{*}{310} & \multirow{1}{*}{1.73} \\
\hline
\textbf{KOI 1375.01} & \multirow{1}{*}{295} & \multirow{1}{*}{1.78} & \multirow{1}{*}{300} & \multirow{1}{*}{1.77} & \multirow{1}{*}{302} & \multirow{1}{*}{1.76} & \multirow{1}{*}{304} & \multirow{1}{*}{1.76} & \multirow{1}{*}{306} & \multirow{1}{*}{1.75} & \multirow{1}{*}{308} & \multirow{1}{*}{1.75} & \multirow{1}{*}{309} & \multirow{1}{*}{1.74} \\
\hline
\textbf{\em{KOI 1377.01}} & \multirow{1}{*}{297} & \multirow{1}{*}{1.78} & \multirow{1}{*}{599} & \multirow{1}{*}{1.77} & \multirow{1}{*}{603} & \multirow{1}{*}{1.76} & \multirow{1}{*}{600} & \multirow{1}{*}{1.75} & \multirow{1}{*}{611} & \multirow{1}{*}{1.75} & \multirow{1}{*}{616} & \multirow{1}{*}{1.74} & \multirow{1}{*}{620} & \multirow{1}{*}{1.74} \\
\hline
\textbf{\em{KOI 1379.01}} & \multirow{1}{*}{297} & \multirow{1}{*}{1.78} & \multirow{1}{*}{599} & \multirow{1}{*}{1.77} & \multirow{1}{*}{603} & \multirow{1}{*}{1.76} & \multirow{1}{*}{600} & \multirow{1}{*}{1.75} & \multirow{1}{*}{611} & \multirow{1}{*}{1.75} & \multirow{1}{*}{616} & \multirow{1}{*}{1.74} & \multirow{1}{*}{620} & \multirow{1}{*}{1.74} \\
\hline
\textbf{KOI 1423.01} & \multirow{1}{*}{ --} & \multirow{1}{*}{ -- } & \multirow{1}{*}{ --} & \multirow{1}{*}{ -- } & \multirow{1}{*}{ --} & \multirow{1}{*}{ -- } & \multirow{1}{*}{300} & \multirow{1}{*}{1.76} & \multirow{1}{*}{303} & \multirow{1}{*}{1.75} & \multirow{1}{*}{ --} & \multirow{1}{*}{ -- } & \multirow{1}{*}{ --} & \multirow{1}{*}{ -- } \\
\hline
\textbf{KOI 1426.01} & \multirow{3}{*}{ --} & \multirow{3}{*}{ -- } & \multirow{3}{*}{300} & \multirow{3}{*}{1.77} & \multirow{3}{*}{302} & \multirow{3}{*}{1.76} & \multirow{3}{*}{304} & \multirow{3}{*}{1.76} & \multirow{3}{*}{306} & \multirow{3}{*}{1.75} & \multirow{3}{*}{308} & \multirow{3}{*}{1.75} & \multirow{3}{*}{310} & \multirow{3}{*}{1.74} \\
\textbf{KOI 1426.02} & &  &  &  &  &  &  &  &  &  &  &  &  &  \\
\textbf{KOI 1426.03} & &  &  &  &  &  &  &  &  &  &  &  &  &  \\
\hline
\textbf{KOI 1429.01} & \multirow{1}{*}{ --} & \multirow{1}{*}{ -- } & \multirow{1}{*}{ --} & \multirow{1}{*}{ -- } & \multirow{1}{*}{ --} & \multirow{1}{*}{ -- } & \multirow{1}{*}{297} & \multirow{1}{*}{1.78} & \multirow{1}{*}{303} & \multirow{1}{*}{1.75} & \multirow{1}{*}{305} & \multirow{1}{*}{1.75} & \multirow{1}{*}{307} & \multirow{1}{*}{1.74} \\
\hline
\textbf{KOI 1463.01} & \multirow{1}{*}{297} & \multirow{1}{*}{1.78} & \multirow{1}{*}{301} & \multirow{1}{*}{1.77} & \multirow{1}{*}{303} & \multirow{1}{*}{1.76} & \multirow{1}{*}{305} & \multirow{1}{*}{1.75} & \multirow{1}{*}{307} & \multirow{1}{*}{1.75} & \multirow{1}{*}{309} & \multirow{1}{*}{1.74} & \multirow{1}{*}{615} & \multirow{1}{*}{1.74} \\
\hline
\textbf{KOI 1472.01} & \multirow{1}{*}{ --} & \multirow{1}{*}{ -- } & \multirow{1}{*}{ --} & \multirow{1}{*}{ -- } & \multirow{1}{*}{ --} & \multirow{1}{*}{ -- } & \multirow{1}{*}{292} & \multirow{1}{*}{1.78} & \multirow{1}{*}{304} & \multirow{1}{*}{1.75} & \multirow{1}{*}{306} & \multirow{1}{*}{1.74} & \multirow{1}{*}{308} & \multirow{1}{*}{1.73} \\
\hline
\textbf{KOI 1478.01} & \multirow{1}{*}{295} & \multirow{1}{*}{1.78} & \multirow{1}{*}{299} & \multirow{1}{*}{1.77} & \multirow{1}{*}{ --} & \multirow{1}{*}{ -- } & \multirow{1}{*}{ --} & \multirow{1}{*}{ -- } & \multirow{1}{*}{ --} & \multirow{1}{*}{ -- } & \multirow{1}{*}{304} & \multirow{1}{*}{1.74} & \multirow{1}{*}{306} & \multirow{1}{*}{1.74} \\
\hline
\textbf{KOI 1486.01} & \multirow{2}{*}{295} & \multirow{2}{*}{1.78} & \multirow{2}{*}{299} & \multirow{2}{*}{1.77} & \multirow{2}{*}{301} & \multirow{2}{*}{1.76} & \multirow{2}{*}{295} & \multirow{2}{*}{1.78} & \multirow{2}{*}{306} & \multirow{2}{*}{1.75} & \multirow{2}{*}{308} & \multirow{2}{*}{1.75} & \multirow{2}{*}{310} & \multirow{2}{*}{1.74} \\
\textbf{KOI 1486.02} & &  &  &  &  &  &  &  &  &  &  &  &  &  \\
\hline
\textbf{KOI 1503.01} & \multirow{1}{*}{296} & \multirow{1}{*}{1.78} & \multirow{1}{*}{300} & \multirow{1}{*}{1.77} & \multirow{1}{*}{302} & \multirow{1}{*}{1.76} & \multirow{1}{*}{304} & \multirow{1}{*}{1.76} & \multirow{1}{*}{306} & \multirow{1}{*}{1.75} & \multirow{1}{*}{308} & \multirow{1}{*}{1.75} & \multirow{1}{*}{310} & \multirow{1}{*}{1.74} \\
\hline
\textbf{\em{KOI 1508.01}} & \multirow{1}{*}{ --} & \multirow{1}{*}{ -- } & \multirow{1}{*}{ --} & \multirow{1}{*}{ -- } & \multirow{1}{*}{ --} & \multirow{1}{*}{ -- } & \multirow{1}{*}{292} & \multirow{1}{*}{1.78} & \multirow{1}{*}{304} & \multirow{1}{*}{1.75} & \multirow{1}{*}{306} & \multirow{1}{*}{1.74} & \multirow{1}{*}{308} & \multirow{1}{*}{1.73} \\
\hline
\textbf{KOI 1527.01} & \multirow{1}{*}{295} & \multirow{1}{*}{1.78} & \multirow{1}{*}{299} & \multirow{1}{*}{1.77} & \multirow{1}{*}{301} & \multirow{1}{*}{1.76} & \multirow{1}{*}{303} & \multirow{1}{*}{1.76} & \multirow{1}{*}{305} & \multirow{1}{*}{1.75} & \multirow{1}{*}{308} & \multirow{1}{*}{1.75} & \multirow{1}{*}{309} & \multirow{1}{*}{1.74} \\
\hline
\textbf{\em{KOI 1528.01}} & \multirow{1}{*}{ --} & \multirow{1}{*}{ -- } & \multirow{1}{*}{ --} & \multirow{1}{*}{ -- } & \multirow{1}{*}{ --} & \multirow{1}{*}{ -- } & \multirow{1}{*}{292} & \multirow{1}{*}{1.78} & \multirow{1}{*}{304} & \multirow{1}{*}{1.75} & \multirow{1}{*}{306} & \multirow{1}{*}{1.74} & \multirow{1}{*}{308} & \multirow{1}{*}{1.73} \\
\hline
\textbf{KOI 1535.01} & \multirow{1}{*}{ --} & \multirow{1}{*}{ -- } & \multirow{1}{*}{300} & \multirow{1}{*}{1.76} & \multirow{1}{*}{ --} & \multirow{1}{*}{ -- } & \multirow{1}{*}{303} & \multirow{1}{*}{1.75} & \multirow{1}{*}{305} & \multirow{1}{*}{1.75} & \multirow{1}{*}{307} & \multirow{1}{*}{1.74} & \multirow{1}{*}{309} & \multirow{1}{*}{1.74} \\
\hline
\textbf{\em{KOI 1561.01}} & \multirow{1}{*}{295} & \multirow{1}{*}{1.78} & \multirow{1}{*}{300} & \multirow{1}{*}{1.77} & \multirow{1}{*}{302} & \multirow{1}{*}{1.76} & \multirow{1}{*}{304} & \multirow{1}{*}{1.75} & \multirow{1}{*}{306} & \multirow{1}{*}{1.75} & \multirow{1}{*}{308} & \multirow{1}{*}{1.74} & \multirow{1}{*}{310} & \multirow{1}{*}{1.74} \\
\hline
\textbf{KOI 1564.01} & \multirow{1}{*}{296} & \multirow{1}{*}{1.78} & \multirow{1}{*}{300} & \multirow{1}{*}{1.77} & \multirow{1}{*}{302} & \multirow{1}{*}{1.76} & \multirow{1}{*}{295} & \multirow{1}{*}{1.76} & \multirow{1}{*}{306} & \multirow{1}{*}{1.75} & \multirow{1}{*}{308} & \multirow{1}{*}{1.74} & \multirow{1}{*}{ --} & \multirow{1}{*}{ -- } \\
\hline
\textbf{KOI 1574.01} & \multirow{1}{*}{294} & \multirow{1}{*}{1.78} & \multirow{1}{*}{298} & \multirow{1}{*}{1.77} & \multirow{1}{*}{300} & \multirow{1}{*}{1.76} & \multirow{1}{*}{303} & \multirow{1}{*}{1.76} & \multirow{1}{*}{305} & \multirow{1}{*}{1.75} & \multirow{1}{*}{307} & \multirow{1}{*}{1.75} & \multirow{1}{*}{309} & \multirow{1}{*}{1.74} \\
\hline
\textbf{KOI 1582.01} & \multirow{1}{*}{295} & \multirow{1}{*}{1.78} & \multirow{1}{*}{299} & \multirow{1}{*}{1.77} & \multirow{1}{*}{ --} & \multirow{1}{*}{ -- } & \multirow{1}{*}{ --} & \multirow{1}{*}{ -- } & \multirow{1}{*}{ --} & \multirow{1}{*}{ -- } & \multirow{1}{*}{ --} & \multirow{1}{*}{ -- } & \multirow{1}{*}{ --} & \multirow{1}{*}{ -- } \\
\hline
\textbf{KOI 1596.01} & \multirow{2}{*}{295} & \multirow{2}{*}{1.78} & \multirow{2}{*}{ --} & \multirow{2}{*}{ -- } & \multirow{2}{*}{ --} & \multirow{2}{*}{ -- } & \multirow{2}{*}{ --} & \multirow{2}{*}{ -- } & \multirow{2}{*}{ --} & \multirow{2}{*}{ -- } & \multirow{2}{*}{307} & \multirow{2}{*}{1.74} & \multirow{2}{*}{309} & \multirow{2}{*}{1.74} \\
\textbf{KOI 1596.02} & &  &  &  &  &  &  &  &  &  &  &  &  &  \\
\hline
\textbf{KOI 1598.01} & \multirow{3}{*}{293} & \multirow{3}{*}{1.78} & \multirow{3}{*}{297} & \multirow{3}{*}{1.77} & \multirow{3}{*}{299} & \multirow{3}{*}{1.76} & \multirow{3}{*}{301} & \multirow{3}{*}{1.76} & \multirow{3}{*}{303} & \multirow{3}{*}{1.75} & \multirow{3}{*}{306} & \multirow{3}{*}{1.75} & \multirow{3}{*}{308} & \multirow{3}{*}{1.74} \\
\textbf{KOI 1598.02} & &  &  &  &  &  &  &  &  &  &  &  &  &  \\
\textbf{KOI 1598.03} & &  &  &  &  &  &  &  &  &  &  &  &  &  \\
\hline
\textbf{\em{KOI 1648.01}} & \multirow{1}{*}{ --} & \multirow{1}{*}{ -- } & \multirow{1}{*}{ --} & \multirow{1}{*}{ -- } & \multirow{1}{*}{ --} & \multirow{1}{*}{ -- } & \multirow{1}{*}{300} & \multirow{1}{*}{1.76} & \multirow{1}{*}{303} & \multirow{1}{*}{1.75} & \multirow{1}{*}{ --} & \multirow{1}{*}{ -- } & \multirow{1}{*}{ --} & \multirow{1}{*}{ -- } \\
\hline
\textbf{\em{KOI 1749.01}} & \multirow{1}{*}{ --} & \multirow{1}{*}{ -- } & \multirow{1}{*}{ --} & \multirow{1}{*}{ -- } & \multirow{1}{*}{300} & \multirow{1}{*}{1.76} & \multirow{1}{*}{287} & \multirow{1}{*}{1.77} & \multirow{1}{*}{304} & \multirow{1}{*}{1.75} & \multirow{1}{*}{306} & \multirow{1}{*}{1.75} & \multirow{1}{*}{308} & \multirow{1}{*}{1.74} \\
\hline
\textbf{\em{KOI 1819.01}} & \multirow{1}{*}{ --} & \multirow{1}{*}{ -- } & \multirow{1}{*}{ --} & \multirow{1}{*}{ -- } & \multirow{1}{*}{302} & \multirow{1}{*}{1.76} & \multirow{1}{*}{304} & \multirow{1}{*}{1.76} & \multirow{1}{*}{306} & \multirow{1}{*}{1.75} & \multirow{1}{*}{308} & \multirow{1}{*}{1.75} & \multirow{1}{*}{310} & \multirow{1}{*}{1.74} \\
\hline
\textbf{\em{KOI 2248.01}} & \multirow{4}{*}{ --} & \multirow{4}{*}{ -- } & \multirow{4}{*}{ --} & \multirow{4}{*}{ -- } & \multirow{4}{*}{ --} & \multirow{4}{*}{ -- } & \multirow{4}{*}{297} & \multirow{4}{*}{1.78} & \multirow{4}{*}{303} & \multirow{4}{*}{1.75} & \multirow{4}{*}{305} & \multirow{4}{*}{1.75} & \multirow{4}{*}{307} & \multirow{4}{*}{1.74} \\
\textbf{\em{KOI 2248.02}} & &  &  &  &  &  &  &  &  &  &  &  &  &  \\
\textbf{\em{KOI 2248.03}} & &  &  &  &  &  &  &  &  &  &  &  &  &  \\
\textbf{\em{KOI 2248.04}} & &  &  &  &  &  &  &  &  &  &  &  &  &  \\
\hline
\textbf{\em{KOI 2418.01}} & \multirow{1}{*}{295} & \multirow{1}{*}{1.78} & \multirow{1}{*}{ --} & \multirow{1}{*}{ -- } & \multirow{1}{*}{ --} & \multirow{1}{*}{ -- } & \multirow{1}{*}{ --} & \multirow{1}{*}{ -- } & \multirow{1}{*}{ --} & \multirow{1}{*}{ -- } & \multirow{1}{*}{307} & \multirow{1}{*}{1.74} & \multirow{1}{*}{309} & \multirow{1}{*}{1.74} \\
\hline
\textbf{\em{KOI 2493.01}} & \multirow{1}{*}{294} & \multirow{1}{*}{1.78} & \multirow{1}{*}{298} & \multirow{1}{*}{1.77} & \multirow{1}{*}{300} & \multirow{1}{*}{1.76} & \multirow{1}{*}{303} & \multirow{1}{*}{1.76} & \multirow{1}{*}{305} & \multirow{1}{*}{1.75} & \multirow{1}{*}{307} & \multirow{1}{*}{1.75} & \multirow{1}{*}{309} & \multirow{1}{*}{1.74} \\
\hline
\textbf{KOI 2534.01} & \multirow{2}{*}{297} & \multirow{2}{*}{1.78} & \multirow{2}{*}{302} & \multirow{2}{*}{1.76} & \multirow{2}{*}{304} & \multirow{2}{*}{1.76} & \multirow{2}{*}{296} & \multirow{2}{*}{1.78} & \multirow{2}{*}{308} & \multirow{2}{*}{1.75} & \multirow{2}{*}{310} & \multirow{2}{*}{1.74} & \multirow{2}{*}{312} & \multirow{2}{*}{1.73} \\
\textbf{KOI 2534.02} & &  &  &  &  &  &  &  &  &  &  &  &  &  \\
\hline

\hline

\end{longtable}
\label{tab:obs}

\end{landscape}

\clearpage

\end{document}